\documentclass[pra, 10pt, twocolumn, aps, superscriptaddress, showpacs, groupaddress]{revtex4-2}
\usepackage[T1]{fontenc}
\usepackage{bigints}
\usepackage{amssymb}
\usepackage{graphicx}
\usepackage{amsmath}
\usepackage{dcolumn}
\usepackage{multirow}
\usepackage{hyperref}
\usepackage[normalem]{ulem}
\usepackage{physics}
\usepackage{bm}
\usepackage{comment}
\usepackage{float}
\usepackage[dvipsnames]{xcolor}
\usepackage[caption=false]{subfig}
\usepackage{siunitx}
\usepackage{dsfont}
\usepackage{bbold}
\usepackage[toc]{appendix}
\usepackage{relsize}
\usepackage{leftidx}
\usepackage{bbm}
\usepackage{stackengine}
\usepackage{orcidlink}

\definecolor{lapislazuli}{rgb}{0.15, 0.38, 0.61}
\definecolor{YKblue}{rgb}{0.0, 0.18, 0.65}
\definecolor{carmine}{rgb}{0.81, 0.09, 0.13}
\definecolor{lavender}{rgb}{0.84, 0.79, 0.87}

\hypersetup{
	colorlinks=true,
	linkcolor=YKblue,
	linktoc=page,
	citecolor=blue,
	urlcolor=YKblue
}

\begin{document}
	
\title{Quantum kinetic theory of light-matter interactions in degenerate plasmas}

	\author{J. L. Figueiredo \orcidlink{0000-0001-6090-6400}}
	\email{jose.luis.figueiredo@tecnico.ulisboa.pt} 
    \affiliation{GoLP - Instituto de Plasmas e Fus\~{a}o Nuclear, Instituto Superior T\'{e}cnico, Universidade de Lisboa, Av. Rovisco Pais 1, 1049-001 Lisboa, Portugal}
	
    \author{J. T. Mendon\c{c}a \orcidlink{0000-0001-8951-3395}}
	\affiliation{GoLP - Instituto de Plasmas e Fus\~{a}o Nuclear, Instituto Superior T\'{e}cnico, Universidade de Lisboa, Av. Rovisco Pais 1, 1049-001 Lisboa, Portugal}
	
    \author{H. Ter\c{c}as \orcidlink{0000-0003-2826-4377}}
	\email{hugo.tercas@tecnico.ulisboa.pt}
    \affiliation{GoLP - Instituto de Plasmas e Fus\~{a}o Nuclear, Instituto Superior T\'{e}cnico, Universidade de Lisboa, Av. Rovisco Pais 1, 1049-001 Lisboa, Portugal}

\begin{abstract}
A rigorous treatment of light-matter interactions typically requires an interacting quantum field theory. However, most applications of interest are handled using classical or semiclassical models, which are valid only when quantum-field fluctuations can be neglected. This approximation breaks down in scenarios involving large light intensities or degenerate matter, where additional quantum effects become significant. In this work, we address these limitations by developing a quantum kinetic framework that treats both light and matter fields on equal footing, naturally incorporating both linear and nonlinear interactions. To accurately account for light fluctuations, we introduce a photon distribution function that, together with the classical electromagnetic fields, provides a better description of the photon fluid. From this formalism, we derive kinetic equations from first principles that recover classical electrodynamical results while revealing couplings that are absent in the corresponding classical theory.  Furthermore, by addressing the Coulomb interaction in the Hartree-Fock approximation, we include the role of fermionic exchange exactly in both kinetic and fluid regimes through a generalized Fock potential. The latter provides corrections not only to the electrostatic forces but also to the plasma velocity fields, which become significant in degenerate conditions.

\end{abstract}

\maketitle

\section{Introduction}
The kinetic description of photons in charged media has garnered significant interest due to its profound implications in various advanced fields such as laser-plasma interaction \cite{Young_1997a, Salamin_2006, livro_2013, Zhang_2023, Michel_2023}, nonlinear optics \cite{Joshi_1990, Umstadter_98, Marklund2005}, photon acceleration \cite{Mendonca_1994, Mendanha_1997, Bu_2015}, and, more recently, photon condensation \cite{plasmaBEC1}. In laser-plasma interactions, the coupling between intense laser fields and plasma particles can lead to a variety of instabilities that are crucial  for inertia confinement fusion and advanced particle accelerators \cite{Tajima_1979, Kruer2003, Ohkubo_2007, Picksley_2023, Shukla2011}. These instabilities, including stimulated Raman and Brillouin scattering, arise from the nonlinear response of the plasma to the electromagnetic fields and can significantly affect the efficiency and stability of energy-transfer processes \cite{Shukla2011, Malkin1999}. Understanding these interactions at a kinetic level is essential for predicting and mitigating such instabilities, thereby enhancing the performance of laser-driven plasma applications.


The problem of photon acceleration also constitutes a paradigmatic example in which a kinetic framework comprising the dynamics of both light and matter is essential. In general terms, photon acceleration refers to the process by which photons gains or losses energy through interactions with moving charged waves or particles without being absorbed \cite{Tajima1979, Mendonca1995, Mendonca1999, Mendonca2000}. This phenomenon is not only an important aspect of plasma physics but has also practical implications for generating high-energy photon sources \cite{Tajima1979, Malkin1999}. Moreover, effects such as harmonic generation, self-focusing, and modulational instability, are driven by the nonlinear interactions between light and matter, impacting applications ranging from laser fusion to high-intensity laser-matter interactions \cite{Shukla2011, Rosenbluth1972}.\par

Another important application of photon kinetics is related to photon condensation, occurring under conditions where photons can thermalize and reach a macroscopic quantum state \cite{Klaers2010}. This process is usually addressed by resorting to the Zel'dovich effect, which involves the transfer of energy from the bulk motion of the electrons to the radiation field via Compton scattering \cite{Zeldovich1969}. The kinetic treatment of photon condensation can, under certain conditions, be formulated with the Kompaneets equation, which governs the diffusion of photons in energy space due to comptonization \cite{Kompaneets1957} and describes a phase transition into a condensed photon phase \cite{plasmaBEC1}. However, the question of including the  dynamics of the plasma and understanding how it modifies the photon condensate remains elusive, since it requires a general kinetic formalism capable of treating both matter and light degrees of freedom on equal footing. This requirement is also important for understanding warm dense matter (WDM) \cite{Dornheim2018}, where extreme conditions$-$such as high temperatures and densities far exceeding those of solid-state systems$-$are encountered. The WDM regime is present in various astrophysical environments, including the interiors of giant planets \cite{Vorberger2007, Militzer2008, Wilson2010, Pstow2016}, brown and white dwarfs, and neutron star crusts \cite{Chabrier2000, Glenzer2016, Militzer2021, Daligault2009, Steiner2016}, as well as in experiments involving high-intensity laser interactions with solids \cite{NIF2024, Pak2024, Hurricane2024, Riley2021, Hyland2021}. In these conditions, exchange effects become dominant \cite{frontiers2014}, making classical kinetic theories inadequate.\par\par

While the examples of physical systems described above put in evidence a vast literature on photon kinetics within classical and semi-classical approximations \cite{McDonald1988, Mendonca2000}, a full quantum treatment of the many-body problem is still lacking. It is fair to state that photon quantization is very well understood in the communities of quantum optics, in which matter may or may not be quantized \cite{CohenTannoudji1994}, as in cavity QED realizations \cite{Lechner_2023}. But in plasmas a difficulty arises already at the establishment of a comprehensive description of quantum plasmas \cite{Manfred1_2001, Haas2011}, i.e. a system in which matter degeneracy is properly taken into account, as well as the quantum nature of light-matter interactions. A representative part of the community investigating quantum effects in plasmas  employ quantum hydrodynamics models \cite{Haas2011}, in which the Bohm diffusion is accounted for and the electron Fermi pressure is phenomenologically introduced for closure \cite{Ali2011}. Alternative kinetic formulations of the problem have also been taken into account via a Wigner-Moyal formalism \cite{Bonitz2016}, and considerations about exchange effects have recently been discussed (see Ref. \cite{Haas_2024} and references therein for a recent discussion). However, these results do not offer a clear representation of fermionic degeneracy that can be interpreted using semiclassical concepts.\par

In general terms, situations in which the combination of nonlinear light-matter interactions and quantum degeneracy have an important role are also difficult to describe within the existing semi-classical models. In the language of field theory, photon-photon correlations are described by doublet photon correlators, which can only be related to the classical electromagnetic fields if the state of radiation is sufficiently coherent, or in other words, when quantum fluctuations are small. Since nonlinear interactions typically spoil light coherence, a description in terms of classical fields frequently leads to several inconsistencies. Indeed, a many-body quantum treatment for interacting light and matter that is capable of including both one-photon and two-photon processes as well as matter degeneracy, while maintaining enough analytical traceability, is still absent. \par 

In this paper, we try to fill this gap by introducing a quantum-kinetic treatment of light-matter interaction in nonrelativistic, degenerate quantum plasmas. We describe the photon sector with a vector potential that accounts for the linear light-matter interactions, $\sim \hat {\bf A}$, plus a photon Wigner function to describe nonlinear processes, $\sim \hat {\bf A}^\dagger \hat {\bf A}$, and treat both as independent variables. The matter fields are described by a set of Wigner functions which retain the relevant quantum features. We show that Compton-scattering events give rise to interacting terms involving matter and radiation Wigner functions which, in turn, yield additional light-matter couplings that can not be extracted from the corresponding classical theory. These couplings play an important role for large field intensities or small quantization volumes. Moreover, in the long wavelength limit, these Compton terms have a clear interpretation in terms of fluid equations, since they give rise to additional forces, pressures and currents. On the other hand, photon absorption and emission events couple the vector potential to matter distributions and give rise to Lorentz dynamics, with the lowest-order $\hbar$-contributions corresponding to the Lorentz force. Therefore, apart from recovering all classical electrodynamical effects, we also include the quantum corrections up to second order in light-matter coupling constants. Additionally, we describe Coulomb interactions at the Hartree-Fock level and determine the exact role of exchange energy by deriving a generalized Fock potential and studying its impacts at several levels of approximation. This is only possible because a second-quantized Coulomb interaction was used. By including both higher-order light-matter interactions as well as Coulomb exchange effects, our model is able to describe a wide range of physical scenarios, ranging from solid-state to dense astrophysical plasmas and warm-dense matter. \par 
This paper is organized as follows. In Sec.~II, we provide a comprehensive revision of the many-body quantum problem and discuss the unified quantum-kinetic approach used for both fermionic and bosonic fields. In Sec.~III, we derive the coupled quantum kinetic equations governing the boson and fermion species. In Sec.~IV, we perform a detailed analysis of the effects of the electron exchange (Fock) in the kinetic equations. A fluid model for the photon$+$matter system is obtained in Sec.~V, which allows us to investigate the normal modes of the system in the hydrodynamic regime in Sec.~VI. Finally, a discussion of the main results and some conclusions regarding future applications is given in Sec.~VII. 

\section{Basic formulation}

The quantum field theory of interacting light and matter follows from minimal coupling the radiation fields to the momentum of fermionic charges. Here, we consider a nonrelativistic plasma composed of charged fermions of type $\alpha$, (in the most usual situations $\alpha={e,i}$ with $e$ and $i$ denoting electron and ions, respectively, or $\alpha={e}$ for the case of a degenerate Fermi gas), mass $m_\alpha$ and charge $Q_\alpha$ represented by (spinless) quantum-field operators $\hat \Psi_\alpha(\mathbf r,t)$. The latter verify fermionic commutation relations $\{\hat\Psi_\alpha(\mathbf r,t),\hat\Psi_{\alpha}^\dagger(\mathbf r',t)\} = \delta(\mathbf r - \mathbf  r')$, with $\{\,,\,\}$ being the anti-commutator, whereas if $\alpha\neq \alpha'$ we have $[\hat\Psi_\alpha(\mathbf r,t),\hat\Psi_{\alpha'}^\dagger(\mathbf r',t)] = 0$, with $[\,,\,]$ the commutator. On the other hand, the radiation quantum-field operator is a real vector field denoted by $\hat{\mathbf A}(\mathbf r,t) = \hat{\mathbf{A}}^{(+)}(\mathbf r,t) + \hat{\mathbf{A}}^{(-)}(\mathbf r,t)$, where $ \hat{\mathbf{A}}^{(\pm)}$ are the positive and negative frequency components related by conjugation. The expectation value of the radiation field plays the role of the classical vector potential $\mathbf A(\mathbf r,t) \equiv \big\langle  \hat{\mathbf{A}}(\mathbf r,t)\big\rangle$, where $\langle . \rangle$ corresponds to the expectation value defined below in Eq.~\eqref{exopValue}. This classical field verifies Maxwell's equations, hence forming the basis for describing all classical electrodynamics phenomena. Since $\hat{\mathbf A}(\mathbf r,t)$ represents spin--1 particles, the commutation relations read $[\hat{\mathbf A}(\mathbf r,t),\partial_t \hat{\mathbf A}(\mathbf r',t)] = i\hbar\delta^\perp(\mathbf r -\mathbf r')/\epsilon_0$, with $\epsilon_0$ the vacuum permittivity and $\delta^\perp(\mathbf r )$ the transverse delta function \cite{Cohen1989}. \par 

\subsection{Field Hamiltonian}
The starting point of our discussion is the low-energy QED Hamiltonian representing the total energy of the interacting system. Upon neglecting relativistic corrections, it can be written as 
\begin{widetext}
\begin{align}
\hat{H} &= \ \sum_\alpha \frac{1}{2m_\alpha} \int d\mathbf r \ \hat{\Psi}_\alpha^\dagger(\mathbf r) \Big[-i\hbar\bm{\nabla}_\mathbf{r} - Q_\alpha \hat{\mathbf A}(\mathbf r)\Big]^2 \hat{\Psi}_\alpha(\mathbf r) + \frac{1}{2}\sum_{\alpha,\alpha'}\int d\mathbf r \int d\mathbf r' \ \hat{\Psi}_\alpha^\dagger(\mathbf r) \hat{\Psi}_{\alpha'}^\dagger(\mathbf r') U^{\alpha,\alpha'}(\mathbf r - \mathbf r') \hat{\Psi}_{\alpha'}(\mathbf r') \hat{\Psi}_\alpha(\mathbf r) \nonumber \\
&+ \frac{1}{2}\int d\mathbf r \Bigg[ \epsilon_0 \hat{\mathbf  E}(\mathbf  r)\cdot \hat{\mathbf  E}(\mathbf  r) + \frac{1}{\mu_0}\hat{\mathbf  B}(\mathbf  r) \cdot \hat{\mathbf  B}(\mathbf  r)\Bigg], \label{QEDHam}
\end{align}
\end{widetext}
where $U^{\alpha,\alpha'}(\mathbf r) = Q_\alpha Q_{\alpha'}/(4\pi \epsilon_0 |\mathbf r|)$ is the Coulomb potential, $\mu_0$ is the vacuum permeability, $\hat{\mathbf  E} = -\partial_t \hat{\mathbf  A}$ and $\hat{\mathbf  B} = \bm{\nabla}\times \hat{\mathbf  A}$ denote the electric and magnetic field operators, respectively. Expectation values of the latter define the electromagnetic fields of the corresponding classical theory, $\mathbf  E \equiv \langle \hat{\mathbf  E}\rangle$ and $\mathbf  B \equiv \langle \hat{\mathbf  B}\rangle$. \par 
The set of fields $\{\hat \Psi_\alpha\}$ and $\hat{\mathbf{A}}$ are enough for a rigorous quantum-mechanical description of the system. We can interpret the action of $\hat \Psi_\alpha(\mathbf r,t)$ on quantum states as that of annihilating a particle of type $\alpha$ at position $\mathbf r$ and time $t$. Although requiring more involved arguments, a similar interpretation can be attributed to the photon field \cite{Hawton}. \par 
In what follows, it is convenient to expand the fields in terms of creation and annihilation operators. To do that, let $\{\phi_\mathbf{k}(\mathbf r)\}$ denote a complete set of functions labelled by wavevector quantum number $\mathbf k =(k_x,k_y,k_z)$ of the form $\phi_\mathbf{k}(\mathbf r) = e^{i\mathbf k \cdot \mathbf r}/\sqrt{V}$, with $V = L_x  L_y L_z$ being the quantization volume. The set verifies the completeness relations $\int d \mathbf r \, \phi_{\mathbf k}(\mathbf r)\phi_{\mathbf k'}^\ast(\mathbf r) = \delta_{\mathbf k,\mathbf k'}$ and $\sum_{\mathbf k} \phi_{\mathbf k}(\mathbf r)\phi_{\mathbf k}^\ast(\mathbf r') = \delta(\mathbf r-\mathbf r')$, and hence can be used to decompose the field operators as follows:
\begin{align}
	\hat \Psi_\alpha(\mathbf r,t) &= \sum_{\mathbf k} \phi_\mathbf{k}(\mathbf r) \hat c_{\alpha,\mathbf k}(t) , \label{PsiField}\\
	\hat{\mathbf{A}}(\mathbf r,t) &= \sum_{\mathbf k,\mu}   \bm{\mathcal{A}}_{\mathbf k,\mu} \phi_\mathbf{k} \Big(\hat{a}_{\mathbf k,\mu} (t) + \hat{a}^\dagger_{-\mathbf k,\mu} (t) \Big). \label{Afield}
\end{align}
\par 
Above, $\hat c_{\alpha,\mathbf k}$ denotes the annihilation operator for a fermionic excitation of type $\alpha$ of mode $\mathbf k$ ($\hat c_{\alpha,\mathbf k}^\dagger$ is the corresponding creation operator), while $\hat{a}^\dagger_{\mathbf k,\mu}$ and $\hat{a}_{\mathbf k,\mu}$ are, respectively, the creation and annihilation operators of photons of mode $\mathbf k$ and the polarization $\mu$. In this basis, the commutation relations take the form $\{\hat c_{\alpha,\mathbf k}, \hat c_{\alpha,\mathbf k'}^\dagger\} = \delta_{\mathbf  k,\mathbf k'}$ and $[\hat a_{\mathbf k,\mu}, \hat a_{\mathbf k',\mu'}^\dagger] = \delta_{\mathbf  k,\mathbf k'}\delta_{\mu,\mu'}$, while operators associated with different fields all commute with each other. Additionally, $\bm{\mathcal{A}}_{\mathbf k,\mu} = \bm{\varepsilon}_{\mathbf k,\mu}(\hbar/2\epsilon_0 \omega_{\mathbf k})^{1/2}$ are expansion coefficients, $\bm{\varepsilon}_{\mathbf k,\mu}$ are polarization vectors, $\omega_{\mathbf k}=\sqrt{\omega_\text{p}^2 + c^2k^2}$ is the photon frequency associated to mode $\mathbf k$ propagating inside the plasma, and $\omega_\text{p}$ is the (constant) plasma frequency,
\begin{equation}
    \omega_\text{p} = \sqrt{\frac{1}{\epsilon_0V}\sum_\alpha \frac{Q_\alpha^2}{m_\alpha}  \int d\mathbf r \ \langle \hat\Psi_\alpha^\dagger  \hat\Psi_\alpha \rangle }.
\end{equation}
This expansion in terms of dressed photons with nonzero mass due to plasma oscillations can be used because it forms a complete basis as well. Alternatively, one could use a plane-wave expansion, and it is possible to show that the two expansions are related by a Bogoliubov transformation. For the present case, the massive basis is more convenient since we know that photons below the plasma frequency will not propagate due to intense absorption by the charged fluids. On the contrary, using the massive basis produces interacting matrix elements which are all finite due to the energy gap. \par 
For the photon polarization we choose the Coulomb gauge which determines the transverse condition $\bm{\varepsilon}_{\mathbf k,\mu} \cdot \mathbf k= 0$, and we further require that $\bm{\varepsilon}_{\mathbf k,\mu} \cdot \bm{\varepsilon}_{\mathbf k,\mu'}= \delta_{\mu,\mu'}$, where $\mu \in \{1,2\}$ are the two orthogonal polarization states of the photons. Imposing periodic boundary conditions at spatial infinity for all fields results in discretized values of $k_i$ in multiples of $2\pi/L_i$, hence sums are used. When convenient, we shall take the infinite-volume limit $V \rightarrow \infty$ which amounts to replace $(1/V) \sum_{\mathbf k}$ by $1/(2\pi)^3\int d\mathbf k$.\par
Using the occupation-number basis, all states can be decomposed in terms of eigenstates of the number operators. The action of creation and annihilation operators changes the occupation number of the given state by unit (for the multiplicative factors and other details see, e.g, \cite{Fetter}). Additionally, expectation values of operators are computed using the density matrix $\hat \rho = \sum_{\mathbf n,\mathbf m} \rho_{\mathbf n,\mathbf m}\ket{\mathbf n}\bra{\mathbf m}$ after tracing over a complete set of many-body states, 
\begin{equation}
    \langle \hat O \rangle \equiv \text{Tr} \big[\hat O \hat \rho \big] = \sum_{\mathbf n,\mathbf m} \rho_{\mathbf n,\mathbf m} O_{\mathbf m,\mathbf n},   \label{exopValue}
\end{equation}
where $O_{\mathbf m,\mathbf n} = \bra{\mathbf m} \hat O \ket{\mathbf n}$. In the Sch\"{o}dinger picture, the time evolution of $\langle \hat O \rangle$ follows from the von Neumann equation, $\partial_t \hat \rho(t) = \frac{i}{\hbar}[\hat \rho(t),\hat H]$.
\par  
Introducing the expansions of Eqs.~\eqref{PsiField} and \eqref{Afield} into Eq.~\eqref{QEDHam} yields
\begin{equation}
	\hat H = \hat{H}_\text{mat.} + \hat{H}_\text{rad.} + \hat{H}_\text{int.},
\end{equation}
where $\hat{H}_\text{mat.}$ involves matter (fermionic) operators, $\hat{H}_\text{rad.}$ contains only radiation (photonic) operators and the remaining terms account for light-matter interactions. Moreover, $\hat{H}_\text{mat.}$ contains both free and interacting terms, whereas $\hat{H}_\text{rad.}$ corresponds to noninteracting photon modes. The interactions can be further decomposed into $\hat{H}_\text{int.} = \hat{H}_\text{int.}^{(1)} + \hat{H}_\text{int.}^{(2)}$, where each superscript indicates the corresponding number of radiation fields involved. We obtain
\begin{align}
	 \hat H_\text{mat.} &= \sum_{\alpha,\mathbf k} \xi_{\alpha,\mathbf k}\hat c_{\alpha,\mathbf k}^\dagger \hat c_{\alpha,\mathbf k} \nonumber \\
  &+ \frac{1}{2}\sum_{\alpha,\alpha',\mathbf{k},\mathbf{k'},\mathbf{q}} \  U^{\alpha,\alpha'}_{\mathbf q} \hat c_{\alpha,\mathbf k+\mathbf q}^\dagger \hat c^\dagger_{\alpha',\mathbf k' - \mathbf q} \hat c_{\alpha',\mathbf k'}\hat c_{\alpha,\mathbf k} , \label{Hplasma}\\
	 \hat H_\text{rad.} &= \sum_{\mathbf k, \mu} \hbar \omega_{\mathbf k} \hat a^\dagger_{\mathbf k,\mu} \hat a_{\mathbf k,\mu} , \label{Hphoton}\\
	 \hat H_\text{int.}^{(1)} &= \sum_{\alpha,\mathbf k ,\mathbf k',\mu}  M^{\mu,\alpha}_{\mathbf k, \mathbf k'} (\hat a_{\mathbf k,\mu}\hat c_{\alpha,\mathbf k + \mathbf k'}^\dagger \hat c_{\alpha,\mathbf k'} \nonumber \\
  &+ \hat a^\dagger_{\mathbf k,\mu}c_{\alpha,\mathbf k'}^\dagger c_{\alpha,\mathbf k + \mathbf k'}),  \label{Hint1}\\
	 \hat H_\text{int.}^{(2)} &= \sum_{\alpha,\mathbf{k, k',q},\mu,\mu'} D^{\mu,\mu',\alpha}_{\mathbf k, \mathbf k'} \, \hat a_{\mathbf{k},\mu}^\dagger \hat a_{\mathbf{k'},\mu'}\hat c^\dagger_{\alpha,\mathbf{k'+q}} \hat c_{\alpha,\mathbf{k+q}}  .\label{Hint2}
\end{align}
The kinetic energy of $\alpha$th fermions is $\xi_{\alpha,\mathbf k} = \hbar^2\mathbf k^2 /(2m_\alpha)$ and the Coulomb matrix element reads $U^{\alpha,\alpha'}_{\mathbf q} = Q_\alpha Q_{\alpha'}/(\epsilon_0 V\mathbf q^2)$. The linear interactions $\hat H_\text{int.}^{(1)}$ correspond to photon absorption and emission by matter, with amplitude
\begin{equation}
	M^{\mu,\alpha}_{\mathbf k, \mathbf k'} = -\frac{Q_\alpha}{m_\alpha\sqrt{V}}\bm{\mathcal{A}}_{\mathbf k,\mu}\cdot \hbar \mathbf{k}'.
\end{equation}
Moreover, nonlinear interactions $\hat H_\text{int.}^{(2)}$ lead to scattering (or Compton) collisions which conserve the photon number, with
\begin{equation}
	D_\mathbf{k,k'}^{\mu,\mu',\alpha} = \frac{Q_\alpha^2}{m_\alpha V} \bm{\mathcal{A}}_{\mathbf k,\mu}\cdot\bm{\mathcal{A}}_{\mathbf k',\mu'}.
\end{equation}
Additional double emission and absorption processes contained in $\hat H_\text{int.}^{(2)}$ are neglected under the rotating-wave approximation. These contribute with high-correlation corrections that we discard here. \par
The interplay between linear and nonlinear interactions is governed by the ratio $S=\langle \hat H_{\rm int}^{(2)}\rangle/ \langle\hat H_{\rm int}^{(1)}\rangle$. For degenerate systems, the latter can be estimated from Eqs. \eqref{Hint1} and \eqref{Hint2} as $S\sim QA/mv_F$, where $v_F\sim \hbar   n_e^{1/3}/m_e$ is the Fermi velocity, $n_e$ is the electron density and $A$ is the magnitude of the vector potential. As such, situations in which strong nonlinearities and high degeneracies coexist are found in nature for a wide range of parameters, including dense astrophysical environments in contact with sufficiently intense radiation profiles.
\par 
\subsection{Quantum dynamics and correlators}
The exact solution of the many-body dynamics requires finding exact expressions for all field projections as a function of time, or equivalently, for creation and annihilation operators, after solving their coupled equations of motion. The latter can be derived from the Heisenberg equations, 
\begin{align}
	i\hbar \frac{\partial}{\partial t}  \hat c_{\alpha,\mathbf k}  &=  \xi_{\alpha,\mathbf k}  \hat c_{\alpha,\mathbf k} + \sum_{\alpha',\mathbf k',\mathbf q} U_{\mathbf q}^{\alpha,\alpha'} \hat c^\dagger_{\alpha',\mathbf k' } \hat c_{\alpha',\mathbf k'+ \mathbf q} \hat c_{\alpha,\mathbf k - \mathbf q} \nonumber \\
	& + \sum_{\mu,\mathbf k'} M_{\mathbf k',\mathbf k}^{\mu,\alpha} (\hat a_{\mathbf k' ,\mu} + \hat a^\dagger_{-\mathbf k' ,\mu}) \hat c_{\alpha,\mathbf k - \mathbf k'} \nonumber \\
	& +  \sum_{\mu,\mu',\mathbf k',\mathbf q} D_{\mathbf k',\mathbf k - \mathbf q}^{\mu,\mu',\alpha} \hat  a_{\mathbf k',\mu}^\dagger \hat a_{\mathbf k - \mathbf  q,\mu'} \hat  c_{\alpha,\mathbf k' + \mathbf q},\\	
	i\hbar \frac{\partial}{\partial t}   \hat  a_{\mathbf k,\mu}  &= \hbar \omega_{\mathbf k} \hat a_{\mathbf k,\mu} + \sum_{\alpha,\mathbf k'} M^{\mu,\alpha}_{\mathbf k, \mathbf k'} c_{\alpha,\mathbf k'}^\dagger c_{\alpha,\mathbf k + \mathbf k'} \nonumber   \\
	& +  \sum_{\alpha,\mu',\mathbf{k',q}} D^{\mu,\mu',\alpha}_{\mathbf k, \mathbf k'} \hat a_{\mathbf{k'},\mu'} \hat c^\dagger_{\alpha,\mathbf{k'+q}} \hat c_{\alpha,\mathbf{k+q}} . 
\end{align}
Exact solutions are unattainable due to the presence of interactions, hence some approximation scheme must be employed. \par 
Our ultimate goal is to find the dynamics of observables, which correspond to expectation values of hermitian operators. Since any operator can be expanded in products of creation and annihilation operators, the corresponding observables will depend on the expectation values of such products, which are $c-$valued functions known as \textit{correlators}. Describing the system in terms of these functions is equivalent to a description in terms of creation and annihilation operators. However, since in general we have $\langle \hat O_1 \hat O_2 \rangle \neq \langle \hat O_1 \rangle \langle \hat O_2 \rangle$, this means that one single field can possibly give rise to an infinite number of independent correlators. As an example, consider the case of an interacting field coupled through a two-body potential. The structure of the correlator dynamics takes the form
\begin{align*}
\frac{\partial}{\partial t} \langle c^\dagger c \rangle &= \digamma_1 [\langle c^\dagger c  \rangle] + \digamma_2 [\langle c^\dagger c^\dagger c \; c  \rangle], \\
\frac{\partial}{\partial t}  \langle c^\dagger c^\dagger c \; c  \rangle &= \digamma_3 [ \langle c^\dagger c^\dagger c \;c   \rangle] + \digamma_4 [\langle c^\dagger c^\dagger c^\dagger c \; c \; c \rangle], \\
\frac{\partial}{\partial t} \langle c^\dagger c^\dagger c^\dagger c\; c \; c  \rangle &= \digamma_5[\langle c^\dagger c^\dagger c^\dagger c \; c \; c \rangle]  + \digamma_6[\langle c^\dagger c^\dagger c^\dagger c^\dagger c \; c  \;c  \;c  \rangle],\\
& \cdots 
\end{align*}
where each $\digamma_i$ denotes a generic functional that depends on the details of the Hamiltonian. Thus, in the presence of interactions, this hierarchy couples each $N$-body correlator to the next $(N+1)$-body correlator, resulting in an infinite system that is in all equivalent to the Heisenberg equations of motion. In the present case, expectation values of all possible products between different fields must also be included, thereby increasing the complexity of the problem. \par 
When correlations are not too strong, one way of dealing with the infinite chain is by truncating the $N$-body correlators into all possible combinations of lower-order correlators plus quantum fluctuations. This method is known as the cluster expansion \cite{FRICKE1996479,	KIRA2015185, KIRA2014200} and has been successfully applied in many areas of condensed-matter physics such as solid state \cite{KIRA2006155,PhysRevB.46.12587} and quantum optics \cite{PhysRevA.78.022102, eyre1981cluster}. \par 
As an example, consider the fermionic correlator $\langle \hat c_{\mathbf k_1}^\dagger \hat c_{\mathbf k_2}^\dagger \hat c_{\mathbf k_3} \hat c_{\mathbf k_4}\rangle$, which arises due to the Coulomb interaction. We may rewrite it as 
\begin{align}
	\langle \hat c_{\mathbf k_1}^\dagger\hat c_{\mathbf k_2}^\dagger &\hat c_{\mathbf k_3} \hat c_{\mathbf k_4}\rangle = \langle \hat c_{\mathbf k_1}^\dagger \hat c_{\mathbf k_4}\rangle \langle  \hat c_{\mathbf k_2}^\dagger \hat c_{\mathbf k_3} \rangle   \nonumber \\
 & - \langle \hat c_{\mathbf k_1}^\dagger \hat c_{\mathbf k_3}\rangle \langle \hat c_{\mathbf k_2}^\dagger  \hat c_{\mathbf k_4}\rangle+ \delta \langle \hat c_{\mathbf k_1}^\dagger\hat c_{\mathbf k_2}^\dagger \hat c_{\mathbf k_3} \hat c_{\mathbf k_4}\rangle. \label{HFappF}
\end{align}
The two first terms are known as the Hartree-Fock contributions, while the third represents the quantum fluctuations. The latter can be interpret as the departure of the original correlator from its classical value, and becomes more important as the strength of the interaction is increased. Higher-order correlators have similar expansions (see \cite{FRICKE1996479} for a complete discussion). Applying the cluster expansion to order $N$ corresponds to neglecting the fluctuations of $(N+1)$-body correlators, thus discarding all higher-order correlators as well. By doing so, a closed system of equations for a finite number of correlators can be established, which can then be used to approximate the dynamics to any desired order, akin to the BBGKY truncation scheme of classical kinetics \cite{Liboff2003}.  


\subsection{Phase-space description}

After establishing the cluster expansion, the quantum dynamics reduces to an interacting theory involving a finite number of correlators. The arguments of these correlators are the labels of the creation and annihilation operators, which in the present case are, apart from polarization, wavevector quantum numbers. A more convenient set of arguments is desired if one wants to associate these correlators with generalized distribution functions that extend their classical counterparts to the quantum regime. This is possible with the help of the Wigner transform, which provides the basis of quantum kinetic theory \cite{gronewold,moyal}. Since the structure of correlators is different for fermionic and bosonic sectors, below we discuss the two cases separately. 
\subsubsection{Fermionic sector}
When the Hamiltonian conserves the total number of fermions, all fermionic correlators of the form $\langle \hat c_{\mathbf k}\rangle$ must vanish. For this reason, the first nonvanishing fermionic observables take the form 
\begin{equation}
	\big\langle \hat O \big\rangle = \sum_{\mathbf k,\mathbf k'} O_{\mathbf k,\mathbf k'} \langle c^\dagger_{\mathbf k} c_{\mathbf k'} \rangle , \label{obs}
\end{equation}
with $O_{\mathbf k,\mathbf k'}  = \bra{\mathbf k}\hat O \ket{\mathbf k'}$. It can be easily shown that Eq.~\eqref{obs} can be rewritten as
\begin{equation}
	\big\langle \hat O \big\rangle = \frac{1}{V}\sum_{\mathbf k} \int d\mathbf r \ \mathcal W(\mathbf r,\mathbf k,t) \   \mathbb{W}\big[\hat O \big](\mathbf r,\mathbf k) ,\label{OpW}
\end{equation}
where 
\begin{equation}
  \mathbb{W}\big[\hat O \big](\mathbf r,\mathbf k) =  \sum_{\mathbf q} e^{i\mathbf q \cdot \mathbf r} \mathcal O_{\mathbf k - \mathbf q/2,\mathbf k + \mathbf q/2}  
\end{equation}
defines the Wigner transform of $\hat O$ and 
\begin{equation}
\mathcal W(\mathbf r,\mathbf k,t) = \sum_{\mathbf  q} e^{i\mathbf q\cdot \mathbf r} \langle \hat c^\dagger_{\mathbf k - \mathbf q/2} \hat c_{\mathbf k + \mathbf q/2} \rangle \label{WignerF}
\end{equation}
is the Wigner function \cite{wigner}. \par 
Equation~\eqref{OpW} shows that quantum expectation values can be \textit{exactly} retrieved from phase-space integrations involving $c-$valued functions instead of operators. This is analogous to the classical case, provided that the classical distribution is replaced by the Wigner function. Moreover, the equation of motion for $\mathcal W$ approaches the Boltzmann equation when the quantum corrections are neglected, indicating that the Wigner function can indeed be interpreted as the quantum counterpart to the classical distribution. Including the quantum corrections, however, leads to solutions which are not positive definite, preventing us from interpreting the Wigner function as a phase-space density. The negative values signal purely quantum phenomena translated to phase space. \par 
As an example, take the density operator $\hat n(\mathbf x)$, whose Wigner transform reads $ \mathbb{W}\big[\hat n(\mathbf x) \big](\mathbf r,\mathbf k)= \delta(\mathbf r-\mathbf x)$. Using Eq.~\eqref{OpW} we obtain
\begin{equation}
\langle \hat n(\mathbf x) \rangle = \frac{1}{V} \sum_{\mathbf k} \mathcal  W(\mathbf x,\mathbf k,t) , \label{density}
\end{equation}
which is the same as its classical analogue. Other one-body observables of interest, such as current density or local temperature, are determined similarly.\par 
Equivalent expressions for any $N$-body observable with $N>1$ can also be established upon defining higher-order Wigner functions and performing phase-space integrations with these new distributions. We can thus establish a theory for a set of coupled distribution functions that is totally equivalent to the original quantum field theory, with the advantage of maintaining a close similarity with the classical case, which allows for an easier interpretation of the many-body quantum dynamics. \par 
Since a large set of distributions functions is not desired for practical calculations, the cluster expansion introduced before can be applied, leading to a closed theory in terms of distribution functions and fields. We stress that the resulting theory still retains all quantum corrections (i.e., higher-order terms in $\hbar$) but neglects higher-order correlations with respect to the coupling strength. While $\hbar$-corrections express the uncertainty principle in phase-space, quantum fluctuations are related to correlations that can have both a classical or a quantum nature. \par

\subsubsection{Bosonic sector}
For boson fields, correlators of the form $\langle \hat a_{\mathbf k,\mu} \rangle$ are in general not zero, and one-body bosonic observables take the form 
\begin{equation}
	\langle \hat O \rangle = \sum_{\mathbf k, \mu} \Big( \mathcal O_{\mathbf k,\mu} \langle \hat a_{\mathbf k,\mu} \rangle + \mathcal O_{\mathbf k,\mu}^\ast \langle \hat a^\dagger_{\mathbf k,\mu} \rangle \Big) ,
\end{equation}
with $\mathcal O_{\mathbf k,\mu}$ being expansion coefficients related with $\hat O$. These include the classical vector potential as well as the electric and magnetic fields, and are usually taken into account by expressing the bosonic sector in terms of $\mathbf A(\mathbf r,t) = \langle \hat{\mathbf A}(\mathbf r)\rangle$, based on which classical electrodynamics is defined. \par 
From the field-theoretical point of view, classical electrodynamics corresponds to the limit where all quantum fluctuations are neglected, i.e., $\langle \hat a^\dagger_{\mathbf k,\mu} \hat F \rangle \simeq \langle \hat a_{\mathbf k,\mu} \rangle  \langle \hat F \rangle$, $\langle \hat a^\dagger_{\mathbf k,\mu} \hat a_{\mathbf k',\mu'} \hat F \rangle \simeq \langle \hat a^\dagger_{\mathbf k,\mu} \rangle  \langle \hat a_{\mathbf k',\mu'} \rangle  \langle \hat F \rangle$ and so on, where $\hat F$ denotes any fermionic operator. This approximation permits to rewrite all correlators in terms of classical fields and close the theory. While this procedure may be valid in many situations, there are also important cases where photon fluctuations are important. Examples are nanoscale photonics \cite{P1}, semi-conductor microcavities \cite{P2}, high-intensity light-matter interactions \cite{P3,P3.1} or photon condensation \cite{P4,P4.2}. In all these cases, photon fluctuations give rise to corrections to the field intensity which should not be discarded. \par 
The intensity of the photon field is defined from 
\begin{equation}
	I(\mathbf r,t) = \epsilon_0 c\big\langle \hat{\mathbf E}^{(-)}(\mathbf r,t)\cdot \hat{\mathbf E}^{(+)}(\mathbf r,t) \big\rangle, \label{IntensityE0}
\end{equation}
which corresponds to a two-body photon observable, and therefore we can write
\begin{equation}
	I(\mathbf r,t) = \sum_{\mu,\mu'}\sum_{\mathbf k,\mathbf q} \mathcal I_{\mathbf k -\mathbf q/2,\mathbf k+\mathbf q/2}^{\mu,\mu'}(\mathbf r) \langle \hat a^\dagger_{\mathbf k - \mathbf q/2,\mu} \hat a_{\mathbf k+\mathbf q/2,\mu'} \rangle,  \label{IntensityE}
\end{equation}
where $\mathcal I_{\mathbf k,\mathbf k'}^{\mu,\mu'}(\mathbf r)$ are coefficients determined from the quantization modes. In classical electrodynamics, the above expression is approximated by $I(\mathbf r,t) = \epsilon_0 c|\mathbf E(\mathbf r,t)|^2$, which follows from replacing $\langle \hat a^\dagger_{\mathbf k,\mu} \hat a_{\mathbf k',\mu'} \rangle$ by $\langle \hat a^\dagger_{\mathbf k,\mu} \rangle  \langle \hat a_{\mathbf k',\mu'} \rangle$ in the above. When the difference between the latter quantities is large, the field intensity will significantly depart from its classical value, leading to inconsistencies in the classical electrodynamical approach (see the discussion in Appendix \ref{apB}). \par 

To account for intensity fluctuations, the second bosonic correlator must be retained. We can do that by defining a photon Wigner function akin to the fermionic case. Due to the polarization degrees of freedom, this function is generalized to an hermitian matrix defined in polarization space as $ \mathcal N_{\mu,\mu'} = \sum_{\mathbf  q} e^{i\mathbf q\cdot \mathbf r} \langle \hat a^\dagger_{\mathbf k - \mathbf q/2,\mu} \hat a_{\mathbf k + \mathbf q/2,\mu'} \rangle $. It is now easy to show that Eq.~\eqref{IntensityE0} can be \textit{exactly} rewritten in terms of the Wigner function. When spatial variations are slow compared to the relevant wavelength range \footnote{Here, the relevant wavelengths are the range $[\lambda_0 - \sigma_{\lambda}, \lambda_0 + \sigma_{\lambda} ]$, where $\lambda_0$ is the central wavelength and $\sigma_{\lambda}$ the bandwidth. These are defined as $\lambda_0  = \frac{1}{\sum_{\mu, \mathbf k} \mathcal N_{\mu,\mu}} \sum_{\mu, \mathbf k} \frac{2\pi}{|\mathbf k|} \mathcal N_{\mu,\mu}$ and $\sigma_\lambda^2  = \frac{1}{\sum_{\mu, \mathbf k} \mathcal N_{\mu,\mu}} \sum_{\mu, \mathbf k}  (\frac{2\pi}{|\mathbf k|})^2  \mathcal N_{\mu,\mu}-  \lambda_0^2$.}, it follows that $\mathcal I_{\mathbf k -\mathbf q/2,\mathbf k+\mathbf q/2}^{\mu,\mu'}(\mathbf r) \simeq V^{-1} \delta_{\mu,\mu'} \hbar \omega_{\mathbf k}e^{i\mathbf q\cdot \mathbf r}$ and we obtain
\begin{equation}
	I(\mathbf r,t) = \frac{1}{V}  \sum_{\mathbf k,\mu}   \hbar\omega_{\mathbf k} c \,\mathcal N_{\mu,\mu}(\mathbf r,\mathbf k,t),
\end{equation}
which corresponds to the energy-current density of particles with energy $\hbar \omega_{\mathbf k}$ and distribution function $\mathcal N$. Any two-body photon observable, such as the photon density or occupation number, can also be exactly retrieved from $\mathcal N_{\mu,\mu'}$. \par 

\section{Quantum kinetic equations}

In this section we establish the quantum kinetic model that follows from considering the second photon correlator as an independent dynamical variable. Therefore, we shall describe the many-body system using the set of kinetic functions:
\begin{align}
\mathbf{A}(\mathbf r,t) &= \sum_{\mathbf k,\mu}   \bm{\mathcal{A}}_{\mathbf k,\mu} \phi_\mathbf{k}(\mathbf r) \langle \hat{a}_{\mathbf k,\mu}\rangle + \text{c.c} ,\label{Afield2}\\
 \mathcal N_{\mu,\mu'}(\mathbf r,\mathbf k,t) &= \sum_{\mathbf  q} e^{i\mathbf q\cdot \mathbf r} \langle \hat a^\dagger_{\mathbf k - \mathbf q/2,\mu} \hat a_{\mathbf k + \mathbf q/2,\mu'} \rangle ,\label{PWignerF2} \\
\mathcal W_{\alpha}(\mathbf r,\mathbf k,t) &= \sum_{\mathbf  q} e^{i\mathbf q\cdot \mathbf r} \langle \hat c^\dagger_{\alpha,\mathbf k - \mathbf q/2} \hat c_{\alpha,\mathbf k + \mathbf q/2} \rangle .\label{plasmaWigner}
\end{align} 
The linear light-matter interactions couple the matter to electromagnetic fields and are thus governed by the vector potential, while the photon Wigner function is used to describe the (nonlinear) processes that couple to the intensity. 
\par 

\subsection{Coupled kinetic equations}
The equation of motion for $\mathbf{A}(\mathbf r,t)$ follows from Maxwell's equation for the field, 
\begin{equation}
		\Big(\frac{\partial^2}{\partial t^2} - c^2\bm{\nabla}_\mathbf{r}^2  \Big)  \mathbf A = \frac{1}{\epsilon_0} \bm{j}, \label{MXWeq}
\end{equation}
and no cluster expansion is required. Here $\bm{j}(\mathbf r,t)$ denotes the total electric current-density, coupling the vector field to matter distributions through
\begin{equation}
    \bm{j}(\mathbf r,t) = \frac{1}{V} \sum_{\alpha,\mathbf k} Q_\alpha\frac{\hbar \mathbf k}{m_\alpha}  \mathcal W_{\alpha}(\mathbf r,\mathbf k,t).
\end{equation} 
Note that the function above corresponds to the expectation value of the electric-current operator, $\widehat{\bm j}(\mathbf r,t) =  \sum_{\alpha} \frac{Q_\alpha}{m_\alpha} \int d\mathbf r \ \hat\Psi^\dagger_\alpha(\mathbf r,t) (\frac{\hbar}{i}\bm{\nabla}_{\mathbf r} )\hat \Psi_\alpha(\mathbf r,t)$, and can thus be related to matter distributions through Eq.~\eqref{OpW}.
\par
Equations of motion for the Wigner functions require a more involved procedure, which is detailed in Appendix~\ref{appW}. The result can be compactly written as
\begin{align}
\frac{\partial}{\partial t}   \mathcal N_{\mu,\mu'} + \mathcal K\big[  \mathcal N_{\mu,\mu'} , \hbar \omega \big] &= \Lambda_{\mu,\mu'}^{(1)} + \Lambda_{\mu,\mu'}^{(2)} \  , \label{photon1}\\
\frac{\partial}{\partial t} \mathcal W_{\alpha} +\mathcal K \big[\mathcal W_{\alpha}, \mathcal E_{\alpha} \big] &= \mathcal C_\alpha^{(1)} + \mathcal C_\alpha^{(2)} .\label{plasmaWigner1} 
\end{align}
The left-hand sides contain the effects in the absence of light-matter interactions, with $\mathcal K$ representing a differential operator defined as
\begin{align}
    &\mathcal K\big[ f(\mathbf r,\mathbf k), g(\mathbf r,\mathbf k) \big] = \nonumber\\
    &  f (\mathbf r,\mathbf k) \frac{2}{\hbar}\sin\left(\frac{1}{2} \overleftarrow{\bm{\nabla}}_\mathbf{r} \cdot \overrightarrow{\bm{\nabla}}_\mathbf{k} - \frac{1}{2}  \overleftarrow{\bm{\nabla}}_\mathbf{k} \cdot \overrightarrow{\bm{\nabla}}_\mathbf{r} \right)g(\mathbf r,\mathbf k). 
\end{align} 
In Eq.~\eqref{photon1}, $\hbar\omega \equiv \hbar\omega_{\mathbf k}$ are the photon-mode energies with no spatial dependence, hence the second term in the argument above vanishes. This can be easily deduced from the Taylor expansion of $\mathcal K$. On the other hand, $\mathcal E_{\alpha}$ corresponds to the phase-space energy of $\alpha$ particles in the absence of radiation,
\begin{align}
\mathcal E_\alpha(\mathbf r,\mathbf k,t) =  \xi_{\alpha,\mathbf k} + Q_\alpha \Phi^\text{H}(\mathbf r,t)  +\Phi_\alpha^\text{F}(\mathbf r,\mathbf k,t), \label{selfenergy}
\end{align}  
with $\Phi^\text{H}$ and $\Phi^\text{F}_\alpha$ denoting the Hartree and Fock potentials, respectively. Both these functions translate the effect of Coulomb interactions in the Hartree-Fock approximation, and are determined from matter distributions through self-consistent relations. The Hartree potential represents electrostatic effects and reads as a solution to the Poisson equation,
\begin{equation}
   \Phi^\text{H}(\mathbf r,t) = \frac{1}{4\pi \epsilon_0} \int d\mathbf r' \ \frac{\sigma(\mathbf r',t)}{|\mathbf r - \mathbf r'|},
\end{equation}
with $\sigma(\mathbf r,t) = V^{-1} \sum_{\alpha,\mathbf k} Q_\alpha \mathcal W_\alpha(\mathbf r,\mathbf k,t)$ being the charge density expectation value. On the other hand, the Fock potential is a purely quantum contribution that introduces an additional intra-particle coupling that has been neglected in previous kinetic models. It is defined as 
\begin{equation}
	 \Phi^\text{F}_\alpha(\mathbf r,\mathbf k,t) = - \sum_{\mathbf q}  U^{\alpha,\alpha}_{\mathbf q}  \mathcal W_{\alpha} (\mathbf r,\mathbf k  + \mathbf q,t). \label{closureFock}
\end{equation}
Contrarily to $\Phi^\text{H}$, the relation between the Fock potential and each distribution function is local in space and can not be expressed in terms of the density. Instead, it depends on the entire distribution function along the $\mathbf k$ direction and translates the exclusion principle. A more detailed analysis of $\Phi^\text{F}_\alpha$ is left to Section~\ref{SecV}. \par  
The right-hand side of Eqs.~\eqref{plasmaWigner1} and \eqref{photon1} contain the effect of light-matter interactions. For an arbitrary quantum state, each of these terms depends on the kinetic variables through convoluted relations which can be found in Appendix~\ref{appW}. However, when the radiation frequency largely surpasses the typical energy of plasma oscillations (high-frequency limit), simpler light-matter terms can be derived. In particular, only the diagonal elements of $ \mathcal N_{\mu,\mu'}$ need to be considered, so we can take the trace of Eq.~\eqref{photon1} and use the equation for $\mathcal N \equiv \sum_{\mu}  
\mathcal N_{\mu,\mu}$ instead, with redefined light-matter terms $\Lambda^{(i)} \equiv \sum_{\mu} \Lambda^{(i)}_{\mu,\mu}$, reading 
\begin{align}
\Lambda^{(1)} &= -\frac{1}{\hbar} \int d\mathbf r' \ e^{i\mathbf k\cdot\mathbf r'} \bm{j}(\mathbf r_{+})  \overleftrightarrow{\mathcal U}  \mathbf E^{(+)}(\mathbf r_{-})  + \text{c.c}, \label{lambda1} \\
\Lambda^{(2)} & = \frac{1}{\omega_\mathbf{k}} \sum_{\alpha} \Omega_\alpha^2(\mathbf r) \nonumber \\
&\times \sin\left( \frac{1}{2} \overleftarrow{\bm{\nabla}}_{\mathbf k} \cdot  \overrightarrow{\bm{\nabla}}_{\mathbf r} - \frac{1}{2} \overleftarrow{\bm{\nabla}}_{\mathbf r} \cdot  \overrightarrow{\bm{\nabla}}_{\mathbf k}  \right) \mathcal N(\mathbf r,\mathbf k) \label{lambda2}.
\end{align}
Above we defined $\mathbf r_{\pm} = \mathbf r \pm \mathbf r'/2$ and 
\begin{equation}
	\Omega_\alpha(\mathbf r,t) = \sqrt{\frac{Q_\alpha^2 n_\alpha(\mathbf r,t)}{\epsilon_0 m_\alpha}}
\end{equation}
as the local plasma frequency, with $n_\alpha$ being the density of $\alpha$ particles as defined in Eq.~\eqref{density}. Additionally, $\overleftrightarrow{\mathcal U}$ is a matrix operator with polarization indices defined in Eq.~\eqref{Uoperator}. In the same (high-frequency) limit, the plasma collision terms become 
\begin{align}
\mathcal C_\alpha^{(1)}  &= \frac{Q_\alpha}{m_\alpha} \cos \left(\frac{1}{2} \overrightarrow{\bm{\nabla}}_\mathbf{r} \cdot \overrightarrow{\bm{\nabla}}_\mathbf{k} \right) \mathbf  A \cdot \overrightarrow{\bm{\nabla}}_\mathbf{r} \mathcal  W_\alpha \nonumber \\
&-\frac{Q_\alpha  }{m_\alpha} (\mathbf k\cdot \mathbf A )\times  2 \sin \left(\frac{1}{2} \overleftarrow{\bm{\nabla}}_\mathbf{r} \cdot \overrightarrow{\bm{\nabla}}_\mathbf{k} \right) \mathcal W_\alpha, \label{C1_sin}\\
\mathcal C_\alpha^{(2)}  &= \frac{Q_\alpha^2}{\epsilon_0 m_\alpha} \frac{1}{V} \sum_{\mathbf q} \mathcal N(\mathbf r,\mathbf q)\nonumber \\
&\times \sin\left(\frac{1}{2}\overleftarrow{\bm \nabla}_{\mathbf{r}} \cdot \overrightarrow{\bm \nabla}_\mathbf{k}   + \frac{1}{2}\overleftarrow{\bm \nabla}_{\mathbf{r}}\cdot  \overrightarrow{\bm \nabla}_\mathbf{q} \right) \mathcal W_\alpha(\mathbf r,\mathbf k) \frac{1}{\omega_{\mathbf q}}  . 
\label{C2_sin} 
\end{align} \par 
The meaning of these terms should be discussed. First, we note that each term includes an infinite number of differential operators in the form of an $\hbar$-expansion which translates the uncertainty principle and is recurrent in quantum kinetics \cite{gronewold} The first terms of the expansion are the important ones for the classical limit, where kinetic functions vary slowly in space, while the higher-order derivatives become significant in the quantum regime. Upon setting $\hbar = 0$ the classical kinetic theory is recovered. \par 
Equation \eqref{lambda1}, being independent of the photon distribution, represents a source term that contains information about the variation of the number of photons due to light absorption and emission by the charged fluids. It corresponds to the variation of the photon phase-space density per unit time due to the processes contained in Eq.~\eqref{Hint1}. On the other hand, $\Lambda^{(2)}$ translates the effect of light-matter collisions promoted by the Hamiltonian in Eq.~\eqref{Hint2}. Since the latter conserve the photon number, it is natural that, at the kinetic level, light-matter scattering results in a collision term that couples photon and matter distribution functions. A similar interpretation is valid for the matter terms $\mathcal C_\alpha^{(1)}$ and $\mathcal C_\alpha^{(2)}$. \par

\subsection{The semiclassical limit}
In practical calculations, dealing with an infinite number of differential operators is most of the times impossible. Hence, a common approach consists of retaining only the first operators up to a given order in $\hbar$. So, in order to validate and apply our results, in what follows we focus on a simplified theory where only the terms up to first order in spatial derivatives are retained. Typically, retaining the first-order derivatives in a quantum kinetic equation leads to the corresponding semiclassical theory, and therefore the validity of this approach requires moderate fermionic densities and temperatures, as well as slow-varying light fields. To be more specific, the semiclassical limit applied to fermionic equations is valid when the Wigner functions vary in space on a scale that is larger than the corresponding de Broglie wavelength $\lambda_B =  \hbar/\sqrt{2mk_BT}$. This is the case when the quantum-degeneracy parameter relating the electron density $n_e$ with the temperature, $X \equiv 4 \pi^{3/2} \,n_e \, \lambda_B^3$, verifies $X \lesssim 1$ \cite{Manfred1_2001, Haas2011}. For bosons, the same criteria can upon replacing $\lambda_B$ by the corresponding wavelength. The result is a semiclassical model where the classical velocities and forces acquire additional contributions. The latter are associated to the coefficients of $\mathbf r$ and $\mathbf k$ gradients, respectively, akin to the classical case. \par
Applying the semiclassical approximation to the left-hand side of Eq.~\eqref{photon1} yields
\begin{equation}
\mathcal K\big[ \mathcal N, \hbar \omega \big] =  \bm v_\text{ph}(\mathbf k) \cdot \bm{\nabla}_{\mathbf r} \,  \mathcal N,
\end{equation} 
where $\bm v_\text{ph}(\mathbf k) \equiv \bm{\nabla}_{\mathbf k} \omega_{\mathbf k} = c^2 \mathbf k /\omega_{\mathbf k}$ is the photon velocity with no semiclassical correction. On the contrary, for matter particles we get
\begin{align}
\mathcal K\big[ \mathcal W_\alpha , \mathcal E_\alpha \big] &= \left(\bm v_\alpha(\mathbf k) +  \frac{1}{\hbar}\bm{\nabla}_{\mathbf k} \Phi_\alpha^\text{F} \right) \cdot  \bm{\nabla}_{\mathbf r} \mathcal W_\alpha \nonumber \\
&- \Big( Q_\alpha\bm{\nabla}_{\mathbf r} \Phi_\alpha^\text{H} + \bm{\nabla}_{\mathbf r} \Phi_\alpha^\text{F}\Big) \cdot  \frac{1}{\hbar}\bm{\nabla}_{\mathbf k} \mathcal W_\alpha, \label{kinPlasma}
\end{align} 
with $\bm v_\alpha(\mathbf k) = \hbar \mathbf k/m_\alpha$ the fermion velocity. It is important to stress that, while the Hartree term only generates the electrostatic force, the Fock potential provides semiclassical corrections to both the velocity and force. The renormalized fields include the effect of Pauli exclusion and thus have no classical analogue.\par 
The semiclassical approximation applied to the light-matter collision terms provides additional corrections. For photons, we get 
\begin{align}
\Lambda^{(1)} &= -\frac{1}{\hbar \omega_{\mathbf k}} \int d\mathbf r' \ e^{i\mathbf k\cdot\mathbf r'} \bm{j}(\mathbf r_{+})\cdot \mathbf E^{(+)}(\mathbf r_{-})   + \text{c.c}, \\
\Lambda^{(2)} & = -  \mathbf V_\text{ph} \cdot  \bm{\nabla}_{\mathbf r} \, \mathcal N - \mathbf{F}_\text{ph} \cdot  \frac{1}{\hbar}\bm{\nabla}_{ \mathbf k} \,\mathcal N,
\end{align}
where
\begin{align}
\mathbf{F}_\text{ph}(\mathbf r,\mathbf k,t) = \frac{\hbar}{\omega_{\mathbf k}}\bm{\nabla}_{\mathbf r} \sum_{\alpha}  \Omega_\alpha^2 (\mathbf r,t) \,  ,
\end{align}
corresponds to a generalized refractive force and 
\begin{align}
\mathbf{V}_\text{ph}(\mathbf r,\mathbf k,t) = \frac{c^2\mathbf k}{2\omega_{\mathbf k}^3}  \sum_{\alpha} \Omega^2_\alpha(\mathbf r,t)
\end{align}
is a refractive velocity. Both these fields stem from collisions with the plasma, and are hence governed by the fermionic densities. \par 
The semiclassical matter terms reduce to
\begin{align}
	\mathcal C_\alpha^{(1)}  & = \frac{Q_\alpha \mathbf A}{m_\alpha} \cdot \bm{\nabla}_{\mathbf r} \mathcal W_\alpha \nonumber  \\
	&-Q_\alpha\big(  \mathbf E  + \bm v_\alpha  \times \mathbf  B \big) \cdot \frac{1}{\hbar}\bm{\nabla}_{\mathbf k} \mathcal W_\alpha, \\
\mathcal C_\alpha^{(2)}  &= - \tau_\alpha^{-1}  \mathcal   W_\alpha -  \mathbf{F}_\alpha \cdot \frac{1}{\hbar} \bm{\nabla}_{ \mathbf k} \, \mathcal W_\alpha.  \label{c_2}
\end{align}
As expected, linear interactions reduce to the classical Lorentz force (Sec.~\ref{SAOC}) and diamagnetic current. On the other hand, the nonlinear processes give rise to the ponderomotive force
\begin{equation}
   \mathbf{F}_\alpha(\mathbf r,t) = - \frac{ Q_\alpha^2}{2\epsilon_0 m_\alpha}   \bm{\nabla}_{\mathbf r} \ \frac{1}{V} \sum_{\mathbf k} \frac{\hbar }{\omega_{\mathbf k}} \mathcal N(\mathbf r,\mathbf k,t), \label{ponderomotiveF}
\end{equation}
plus a space-dependent decay,
\begin{equation}
    \tau_{\alpha}^{-1}(\mathbf r,t) = - \frac{ Q_\alpha^2}{2\epsilon_0 m_\alpha}  \bm{\nabla}_{\mathbf r}\cdot \frac{1}{V} \sum_{\mathbf k} \frac{c^2 \mathbf k}{\omega_{\mathbf k}^3} \mathcal N(\mathbf r,\mathbf k,t).
\end{equation} 
Note that the expression for $\mathbf{F}_\alpha$ is in accordance with previous results obtained from different methods \cite{ponderomotive1,ponderomotive3}. Moreover it reduces to the well-known formula $\mathbf{F}_\alpha = - Q_\alpha^2 \bm{\nabla} |\mathbf A|^2/m_\alpha$ in the case of classical monochromatic fields.   


\section{A semiclassical fluid model for light and matter}

\label{SecIV}
Next, we focus on the evolution of matter and radiation expectation values defined from distribution functions. In particular, we establish the equations of motion for particle densities and currents using the semiclassical results of the previous section, and arrive to a coupled fluid model with semiclassical corrections. These corrections are related to both the inclusion of exchange Coulomb interactions $-$ leading to extra contributions to the fermionic single-particle velocity, electric force and kinetic pressure $-$, as well as nonlinear light-matter effects related to the microscopic Compton scattering. The latter also contribute to renormalize both fermionic and bosonic fluid currents and pressures, as explained below. \par 
The independent fluid variables are defined as 
\begin{align}
	n_\beta(\mathbf r,t) &= \frac{1}{V} \sum_{\mathbf k} \mathcal W_\beta(\mathbf r,\mathbf k,t),\\
    \mathbf{J}_\beta(\mathbf r,t) &=  \frac{1}{V}\sum_{\mathbf k} \bm v_\beta(\mathbf k) \mathcal W_\beta(\mathbf r,\mathbf k,t),
\end{align}
with $\beta = \alpha,\text{ph}$ and $\mathcal W_\text{ph} \equiv \mathcal N$ denotes the photon distribution. 
\subsection{Matter equations}
From the semiclassical kinetic equations, the following equations for matter quantities are derived:
\begin{align}
&\frac{\partial}{\partial t} n_\alpha+ \bm\nabla\cdot \widetilde{\mathbf{J}}_\alpha  = 0, \label{dtDN} \\
&\frac{\partial}{\partial t} \mathbf{J}_\alpha +\frac{1}{m_\alpha}\bm\nabla \big(P_\alpha + P_\alpha^\text{F}\big)  =  \frac{n_\alpha }{m_\alpha}\widetilde{\mathbf{F}}_\alpha  - \tau_\alpha^{-1} \mathbf{J}_\alpha .\label{dtDJ}
\end{align} 
Above, $ \widetilde{\mathbf{J}}_\alpha \equiv \mathbf{J}_\alpha + \mathbf{J}_\alpha^{(1)} + \mathbf{J}_\alpha^{(2)} + \mathbf{J}_\alpha^\text{F}$ is the total matter current-density, which differs from the bare (kinetic) current $\mathbf{J}_\alpha $ due to the renormalizations stemming from Eqs.~\eqref{kinPlasma} and \eqref{c_2}. Light-matter interactions provide linear,
\begin{align}
	\mathbf{J}_\alpha^{(1)}(\mathbf r,t) &= - \frac{Q_\alpha}{m_\alpha}\mathbf A(\mathbf r ,t)n_\alpha(\mathbf r,t),
\end{align}
and nonlinear renormalizations,
\begin{align}
\mathbf{J}_\alpha^{(2)} (\mathbf r,t) &=  \mathbf V_\alpha(\mathbf r,t) n_\alpha(\mathbf r,t) , \label{J_jp}
\end{align}
with $\mathbf V_\alpha(\mathbf r,t)  = Q_\alpha^2/(2\epsilon_0m_\alpha V)\sum_{\mathbf k} c^2 \mathbf k \mathcal N(\mathbf r,\mathbf k,t)/\omega_{\mathbf k}^3$ denoting a scattering-induced velocity, while exchange-Coulomb interactions result in a degeneracy current of the form
\begin{align}
	\mathbf{J}_\alpha^\text{F} (\mathbf r,t) &= \frac{1}{V}\sum_{\mathbf k} \frac{1}{\hbar} \frac{\partial \Phi^\text{F}}{\partial \mathbf k} \mathcal W_\alpha(\mathbf r,\mathbf k,t). \label{FockJ}
\end{align} 
The latter can be interpreted as an additional current necessary to ensure the exclusion principle. \par 
Moreover, $P_\alpha + P_\alpha^\text{F}$ is the total pressure-tensor which, similarly to the current, is composed by a kinetic contribution, 
\begin{align}
P_\alpha(\mathbf r,t) &=  \frac{1}{V}\sum_{\mathbf k} \frac{\hbar\mathbf k \otimes \hbar\mathbf k }{m_\alpha}  \mathcal W_\alpha(\mathbf r,\mathbf k,t),
\end{align}
plus an exchange contribution,
\begin{align}
P_\alpha^\text{F}(\mathbf r,t) &=  \frac{1}{V} \sum_{\mathbf k} \Big(\frac{\partial \Phi^\text{F}}{\partial \mathbf k} \otimes \mathbf k  \Big) \mathcal W_\alpha(\mathbf r,\mathbf k,t). \label{FockP}
\end{align}
Additionally,
\begin{equation}
	\widetilde{\mathbf{F}}_\alpha =   - Q_\alpha\bm{\nabla}_\mathbf{r} \Phi^\text{H} + \bm{f}_\alpha^\text{F} + Q_\alpha\big(\mathbf{E} + \mathbf{J}_\alpha \times \mathbf{B}\big) + \mathbf F_\alpha 
\end{equation}
is the total force exerted on matter particles. The two first terms steam from the Hartree and Fock potentials and follow from Coulomb interactions with the remaining plasma particles, while the last terms arise from the linear (Lorentz force) and nonlinear (ponderomotive force) couplings to the photon fields. In particular, the exchange force
\begin{equation}
\bm{f}_\alpha^\text{F}(\mathbf r,t) = -\frac{1}{n_\alpha} \frac{1}{V}\sum_{\mathbf k} \mathcal W_\alpha (\mathbf r,\mathbf k,t) \frac{\partial \Phi^\text{F}}{\partial \mathbf r}  \label{FockF}
\end{equation}
represents the Fock correction to the electrostatic force. \par 
\subsection{Photon equations}
Applying the same procedure to the photon quantities yields
\begin{align}
&\frac{\partial}{\partial t} n_\text{ph} + \bm \nabla \cdot \big( \mathbf{J}_\text{ph} - \sum_\alpha  \mathbf{J}_\alpha^{(2)} \big) = S, \label{dtDNp}\\
&\frac{\partial}{\partial t} \mathbf{J}_\text{ph} + \frac{1}{m_\text{ph}}\bm\nabla \Big( P_\text{ph} + \sum_{\alpha} p_\alpha \Big)= \sum_{\alpha}  \mathbf T_\alpha . \label{dtDJp}
\end{align} 
Above, the source term $S$ represents linear interactions with the plasma, while nonlinear interactions reduce to a renormalization of both the photon current and pressure. The latter becomes $P_\text{ph} + \sum_{\alpha} p_\alpha$, containing the usual kinetic contribution
\begin{align}
P_\text{ph}&(\mathbf r,t) =  m_\text{ph} \frac{1}{V} \sum_{\mathbf k}  v_\text{ph}(\mathbf k) \otimes v_\text{ph}(\mathbf k)  \,\mathcal N(\mathbf r,\mathbf k,t), \label{kinP} 
\end{align} 
plus a pressure exerted by Compton collisions with matter,
\begin{align}
p_\alpha (\mathbf r,t) = m_\text{ph}\Omega_\alpha^2(\mathbf r,t) &\frac{1}{V}\sum_{\mathbf k} \frac{1}{2\omega_{\mathbf k}^2} \mathcal N(\mathbf r,\mathbf k,t) \nonumber \\
&\times v_\text{ph}(\mathbf k) \otimes v_\text{ph}(\mathbf k) ,
\end{align} 
verifying $p_\alpha \sim n_\alpha$. Here $m_\text{ph} \equiv \hbar (\partial^2 \omega_{k}/\partial k^2)_{k=0}^{-1}	= \hbar \omega_\text{p}/c^2$ denotes the photon mass. Additionally, 
\begin{align}
	\mathbf T_\alpha(\mathbf  r,t) = &  \frac{1}{V} \sum_{\mathbf k} \frac{1}{2\omega_{\mathbf k}^2} \mathcal N(\mathbf r,\mathbf k,t) \nonumber \\
	& \times \Big(c^2 \mathbb{1} - v_\text{ph}(\mathbf k) \otimes v_\text{ph}(\mathbf k) \Big)\bm{\nabla}\Omega^2_\alpha (\mathbf  r,t)
\end{align}
can be interpreted as a macroscopic force exerted by charge inhomogeneities on the photon fluid. This force arises when spatial variations in the matter density lead to significant changes in the light scattering rates within one photon wavelength. When the typical matter and radiation length scales are sufficiently different, this force vanishes and the kinetics of photons is essentially governed by the pressure. \par 
Photon absorption and emission are described by the source term $S$ which is independent of the photon density. It reads
\begin{equation}
	S(\mathbf r,t) = -\int d\mathbf r' \ \bm{j}(\mathbf r_+) \cdot u(\mathbf r') \mathbf E(\mathbf r_-), \label{SSource}
\end{equation}
where $u(\mathbf r)$ denotes the Fourier transform of the inverse photon dispersion, $u(\mathbf  r) =  \omega_\text{p} K_1(\omega_\text{p}r/c)/(4\pi^2\hbar c^2r)$, with $K_1(x)$ being the modified Bessel function of the second kind. In the limit of an under-dense and over-dense plasma, it reduces to $u(\mathbf  r) = 1/(2\pi^2 \hbar c r^2)$ and  $u(\mathbf r ) = \delta(\mathbf r)/\hbar \omega_\text{p}$, respectively. This semiclassical source is the same as that predicted by Maxwell's equations, which determine that $\partial_t |\mathbf E|^2 \sim \partial_t I_\text{cl.}$ contains a term $\sim \bm j \cdot \mathbf E$ governing the absorption and emission processes (see Appendix \ref{apB}). Quantum corrections to the $\bm j \cdot \mathbf E$ coupling are contained in the remaining terms of $\Lambda^{(1)}$ which are at least second order in spatial derivatives, and therefore are not considered in the semiclassical model. Nonetheless, these corrections will be important at mesoscopic scales. \par 
In fact, Eq.~\eqref{SSource} generalizes the Joule-heating law of absorption of electromagnetic energy from a charged system interacting with a monochromatic light field. The simplified law can be recovered by considering a monochromatic state, $\mathbf E(\mathbf  r,t)\sim e^{i\mathbf q\cdot \mathbf r}$, and integrating the photon continuity equation around a small region with volume $dV\rightarrow 0$, which leads to
\begin{equation}
\frac{\partial}{\partial t} [\hbar \omega_{\mathbf q} n_\text{ph}]  = - \bm j\cdot \mathbf E.
\end{equation}
It is clear that the left-hand side represents the power per unit volume gained by the radiation field. Hence, the power absorbed by the matter is the symmetric,
\begin{equation}
	\frac{d \text{P}}{dV}\Big|_\text{matter} = -\frac{\partial}{\partial t}  [\hbar \omega_{\mathbf q} n_\text{ph}] = \bm{j}\cdot \mathbf E, \label{powerPlasma}
\end{equation}
which corresponds to the Joule effect. \par 

\section{Exchange effects}\label{SecV}
In this section we focus on the impact of exchange interactions at the level of kinetic and fluid equations. We start by estimating the Fock potential of Eq.~\eqref{closureFock} and compare it with the remaining energy scales. This will help to clarify the role of exchange and its importance for different plasma regimes. Then, we calculate the exchange fluid variables for particular cases and show how they can affect the propagation of plasmons at high degeneracies. 

\subsection{Estimating the exchange potential}
Contrary to the Hartree potential, exchange effects associate to each fermion field a different Fock potential $\Phi^\text{F}_\alpha(\mathbf r,\mathbf k,t)$, from which all exchange-fluid variables can be determined. This potential has units of energy and can be interpreted as the Fock counterpart to electrostatic potential. However, a crucial difference between the two is that $\Phi^\text{F}_\alpha$ carries a momentum dependence as well, which, apart from a correction to the force, results in renormalization of the pressure and current, both related to non-vanishing derivatives of $\Phi^\text{F}_\alpha$ with respect to the momentum coordinate [see Eqs.~\eqref{FockJ} and \eqref{FockP}].\par 
While exchange effects are negligible at lower densities or higher temperatures, we expect their role to be important in systems such as solid-state, cold or dense-astrophysical plasmas. In all these cases, including the exchange fluid variables given here for the first time is essential for a rigorous treatment of the plasma.  \par 
To clarify the role of exchange, let us focus on a single type of fermion and assume isotropic equilibrium at a constant temperature $T$ and arbitrary degeneracy. The Wigner function for the latter is the Fermi-Dirac distribution (we drop the index $\alpha$ in the remaining of this section):
\begin{equation}
	\mathcal W_0 (k) =  \frac{g_s}{1 + e^{\lambda_B^2 k^2}e^{-\mu/k_BT}} , \label{W1}
\end{equation} 
where $\lambda_B = \hbar/\sqrt{2mk_BT}$ is the De Broglie length and $g_s = 2$ is the spin degeneracy. The chemical potential $\mu$ is a Lagrange multiplier that fixes the total number of particles $N$ contained in the volume $V$, $N = \sum_{\mathbf k} \mathcal W_0(k)$. For sufficiently large volume, this condition becomes
\begin{equation}
	n = \int \frac{d\mathbf k}{(2\pi)^3} \frac{g_s}{1 + e^{\lambda_B^2 k^2}e^{-\mu/k_BT}},
\end{equation}
which defines a relation between $\mu$, $n$ and $T$. Performing the integration leads to $ X =-\text{Li}_{\frac{3}{2}}\big(-\xi \big)$, where $X = 4 \pi^{3/2} \,n \, \lambda_B^3 $ denotes the degeneracy parameter, $\text{Li}_z(x)$ is the polylogarithm function of order $z$ and $\xi = \text{exp}(\mu/k_BT)$ is the fugacity. Since $\text{Li}_i(x)$ has no analytical inverse, the form of $\mu = \mu(n,T)$ in the entire $n-T$ plane can only be found after numerical inversion.\par 
We proceed to evaluate the exchange potential assuming the Fermi--Dirac equilibrium for the plasma. After some algebra, we can reduce Eq.~\eqref{closureFock} to the following:
\begin{align}
	&\Phi^\text{F}(k) = -\frac{Q^2}{\epsilon_0 2\pi^2}  \frac{1}{\lambda_B} \nonumber \\
	&\int_0^\infty dq \ \int_{-1}^1 dx \ \frac{q^2}{q^2 + \lambda_B^2k^2 -2 k\lambda_B qx} \  \frac{1}{1+\xi^{-1}e^{q^2}}. \label{ExcFD}
\end{align}
The integral over $q$ cannot be performed analytically if one uses the complete form of the Fermi-Dirac distribution. However, analytical approximations can be found by replacing the Fermi-Dirac function by its corresponding low--degeneracy (Maxwell--Boltzmann) or high--degeneracy (Heaviside) limits. In any case, the order of magnitude of $\Phi^\text{F}(k)$ corresponds essentially to its value at $k=0$. For nonzero $k$, one can verify numerically that $\Phi^\text{F}(k) \sim \Phi^\text{F}(0) G(k/k_0)$ with $G(x)$ a function of order unit. Hence, using Eq.~\eqref{ExcFD}, an estimation for the order of magnitude of $\Phi^\text{F}$ is
\begin{equation}
\Phi^\text{F} \sim  \frac{Q^2}{\epsilon_0} \frac{1}{\lambda_B} \text{Li}_{\frac{1}{2}}(-\xi).
\end{equation}
\par 
The value above should be compared with the average kinetic energy of the Fermi gas,
\begin{equation}
	K  = -\frac{k_BT}{n\lambda_B^3} \text{Li}_{\frac{5}{2}}(-\xi),
\end{equation}
by defining the ratio of exchange--to--kinetic energy as
\begin{equation}
	R \equiv \frac{|\Phi^\text{F}|}{K} =  \Bigg(\frac{\lambda_B}{\lambda_D}\Bigg)^2 \  \frac{\text{Li}_{\frac{1}{2}}(-\xi)}{\text{Li}_{\frac{5}{2}}(-\xi)}. \label{ratio}
\end{equation}
The first factor is the ratio between De Broglie and Debye lengths (the latter being $\lambda_D = \sqrt{\epsilon_0 k_B T/nQ^2}$), which is related to the degree of degeneracy. For high temperatures or low densities, this ratio is essentially zero, which corresponds the classical limit. The second factor is a numerical correction imparted by the Fermi--Dirac distribution, which becomes relevant at high degeneracies.\par 

\begin{figure}
\centering 
\includegraphics[scale=0.5]{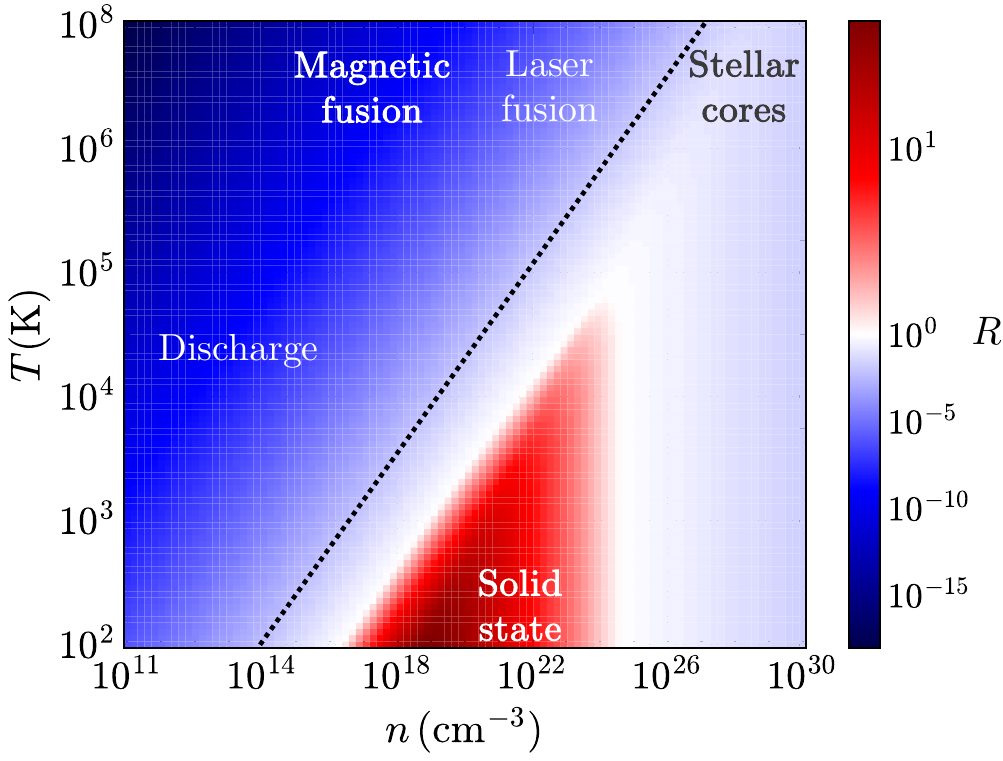}
\caption{(color online) Value of $R$ as a function of density and temperature calculated for an electron gas. Different regimes of interest are highlighted using the reference values of Table.~\ref{tb1}. In the lighter regions, kinetic and exchange effects are comparable ($R \sim 1$). Dark blue regions correspond to classical regimes where exchange effects are negligible, whereas in the red regions exchange effects dominate. The area to the right of the dotted line is defined by $R\geq 0.01$, hence containing the cases in which the exchange potential can be considered non-negligible. These comprise both solid-state as well as dense astrophysical plasmas. }
\label{R_n_T}
\end{figure}

The value of $R$ is fully determined by the density and temperature of the Fermi gas. Different examples are considered in Table~\ref{tb1}, where $R$ is calculated for an electron plasma. Moreover in Fig.~\ref{R_n_T} we depict this ratio for the whole parameter space.\par 

Figure~\ref{R_n_T} shows that classical plasmas, such as fusion or discharge, verify $R\ll 1$ as expected, so the effect of exchange is negligible. On the contrary, for dense astrophysical plasmas such as stellar cores, we find $R\sim 0.05-0.1$. For solid-state plasmas, which are purely quantum, exchange effects dominate. The latter fall within the red region of Fig.~\ref{R_n_T}, where $R$ attains its maximum. Note that, to the right of this region, the degeneracy continues to increase, and so does the exchange potential. However, the rise in kinetic energy is faster such that the overall ratio becomes smaller.
\renewcommand{\arraystretch}{1.2} 
\begin{table}
    \centering
    \begin{tabular}{|>{\centering\arraybackslash}m{2.5cm}|>{\centering\arraybackslash}m{1.4cm}|>{\centering\arraybackslash}m{1cm}|>{\centering\arraybackslash}m{1cm}|>{\centering\arraybackslash}m{1cm}|} \hline
        \textbf{Plasma type} & $n \ (\text{cm}^{-3})$ & $T \ (\text{K})$ &  $X$ &$R$ \\ \hline \hline 
         Magnetic fusion & $10^{15}$ & $10^{7}$ & $10^{-12}$&$10^{-13}$   \\ 
         Discharge & $10^{15}$ & $10^4$ & $10^{-7}$ &$10^{-7}$   \\
         Laser fusion & $10^{23}$ & $10^{7}$ & $10^{-4}$ &$10^{-5}$  \\
      	Stellar cores & $10^{27}$ & $10^7$ & $10$ & $0.1$\\
         Solid--state & $10^{22}$ & $10^2$ & $10^3$ &$10$   \\
         \hline
    \end{tabular}
    \caption{Ratio of exchange--to--kinetic energy for several types of electron plasmas and reference values of density, temperature and fermionic degeneracy. All numbers denote the order of magnitude of the corresponding exact value.} 
    \label{tb1} 
\end{table}

\subsection{Exchange--fluid variables}
Let us now turn our attention to the exchange--fluid variables. In order to find analytical expressions, it is necessary to approximate the Fermi-Dirac distribution for low or high degeneracy conditions, $X \ll 1$ and $X \gg 1$, respectively,
\begin{equation}
	\mathcal  W_0(k) = g_s \times \begin{cases} X e^{-\lambda_B^2 k^2} 
	\quad \quad \quad \,\text{if} \ X \ll 1 , \\
	 \Theta(k_F -k) \quad \quad \ \   \text{if}  \ X \gg 1. \label{FDlimits}
	\end{cases}
\end{equation} \par
\begin{figure}
\centering 
\includegraphics[scale=0.55]{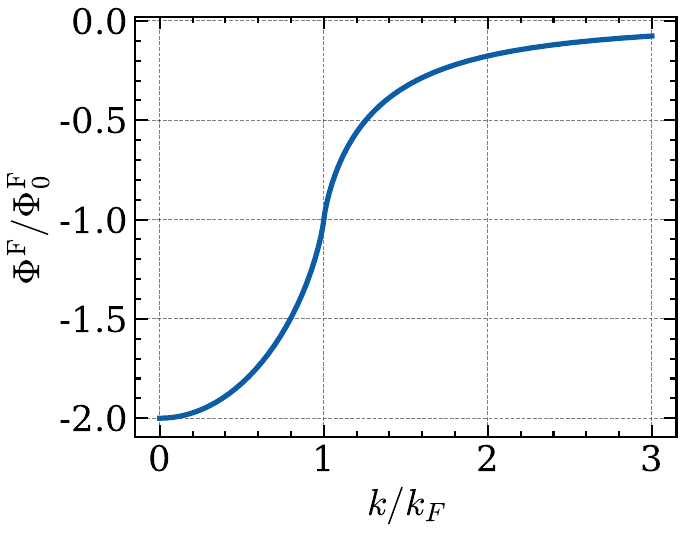}
\caption{(color online) Graphical representation of the normalized exchange potential of a degenerate electron plasma.}
\label{Psi(k)}
\end{figure}

 
Since the role of exchange is relevant only for sufficiently high degeneracies, in what follows we focus on the second case. Using the appropriate Fermi--Dirac limit, we obtain
\begin{equation}
	\Phi^\text{F}(k) = -\Phi^\text{F}_0 G(k_F/k),
\end{equation}
with $\Phi^\text{F}_0 = Q^2 k_F/(2\pi^2 \epsilon_0)$ a constant and
\begin{equation}
	G(x) = 1 + \frac{x^2-1}{2x} \ln(\frac{1+x}{|1-x|}).
\end{equation}
The shape of $\Phi^\text{F}(k)$ is represented in Fig.~\ref{Psi(k)}. \par 
Equations of state relating dependent and independent variables can now be established.  
Since we consider isotropic equilibrium, all currents vanish, while pressures become scalars,
\begin{align}
P &= \frac{ 3^{2/3}\pi^{4/3}\hbar^2 }{5 m} n^{5/3}, \label{EqState1} \\
P^\text{F}&= \frac{3^{1/3} Q^2 }{4\pi^{4/3} \epsilon_0 } n^{4/3}.\label{EqState2} 
\end{align}
Moreover, the exchange force reads
\begin{equation}
\bm{f}^\text{F} = \frac{3^{1/3} Q^2}{ 2 \pi^{4/3} \epsilon_0 n^{2/3}} \bm \nabla n . \label{EqState3} 
\end{equation} \par 
Since exchange effects are usually associated with degeneracy repulsion, it might seem counter-intuitive to find that the exchange force points along the density gradient. However, the repulsive character is contained in the degeneracy pressure $P^\text{F}$, which increases with increasing density as expected. On the other hand, the direction of the force should be such that it leads the system to a state of lower total exchange energy, which, for the particular case, corresponds to a state of higher density. This can be demonstrated by calculating the total exchange energy, which can be found from the exchange potential as $\mathcal E^\text{F}=V^{-1}\sum_{\mathbf k} \int d\mathbf r \  \mathcal W_0 \Phi^\text{F}$, providing
\begin{equation}
    \mathcal E^\text{F} = - g_s \frac{V}{(2\pi)^3} \frac{Q^2}{4\pi\epsilon_0} k_F^4 \sim n^{8/3}. \label{E_F}
\end{equation}
We conclude that $\bm{f}^\text{F}$ has the correct direction (i.e., along $-\bm{\nabla}\mathcal E^\text{F}$). \par 
The result of Eq.~\eqref{E_F} has been known for a long time (see, e.g., Ref.~\cite{Fetter}), despite alternative approaches had been followed, which are typically suited for the ground-state properties only. On the contrary, the method presented here goes beyond previous works since it can be applied to out-of-equilibrium conditions as well, including inhomogeneous and transient regimes. To the best of our knowledge, both the Fock potential and exchange-fluid variables are given here for the first time.

\subsection{Plasmon dispersion at high degeneracies}
Now we focus on the impact of exchange in the propagation of collective modes in highly-degenerate Fermi gases. To do that, let us go back to the fluid model and neglect all light-matter interacting terms. The latter will be considered in the next section. \par 
After expanding each dynamical quantity as an equilibrium value plus fluctuations, $v(\mathbf r,t)=v_0  + \delta v(\mathbf r,t)$ and retaining only first-order fluctuating terms, a linearized model can be established. Moving to Fourier space, we get
\begin{equation}
	\begin{pmatrix}
		-\omega &  q & 0 \\
		v_\text{eff}^2 q  & -\omega  & \frac{n_0Q}{m}q\\
		\frac{Q}{\epsilon_0} & 0 & -\frac{1}{q^2}
	\end{pmatrix} \begin{pmatrix}
		\delta n( q,\omega)\\
		\delta \mathbf J( q,\omega)\\
		\delta \Phi^\text{H}( q,\omega) 
	\end{pmatrix} =0,
\end{equation}
where we defined
\begin{equation}
	v_\text{eff} = \sqrt{\frac{1}{3}v_F^2 - \frac{1}{2}\tilde{v}^2}
\end{equation}
as the \textit{effective plasmon velocity} in degenerate plasmas, with $v_F = \hbar k_F/m$ and $\tilde{v} = \hbar  \omega_\text{p}/(m v_F)$ the Fermi and exchange velocities, respectively. \par
The plasmon dispersion can be obtained by setting the determinant of the dispersive matrix above to zero, leading to
\begin{equation}
	\omega = \sqrt{\omega_\text{p}^2 + v_\text{eff}^2 q^2}, \label{plasmonDisp}
\end{equation}
which, apart from a small modification in the numerical coefficients due to the linearization procedure, is in agreement with previous results obtained from different methods \cite{PhysRev.111.442,PhysRev.121.941,PhysRevE.92.013104}.\par 

We see that different effects contribute to the plasmon dispersion. On the one hand, the plasma frequency represents the contribution of electrostatic interactions, granting the plasmons with an effective mass. On the other hand, the effective velocity contains the effects of kinetic pressure and exchange energy. All these effects become more important as the density increases, however in descending order: $\omega_\text{p} \sim n_0^{1/2}$, $v_F \sim n_0^{1/3}$ and $\tilde{v} \sim n_0^{1/6}$. This explains why exchange effects are more important in solid-state regimes than in dense astrophysical environments, which is in agreement with the results of the previous section (see Fig.~\ref{R_n_T}). \par
\begin{figure}
\centering \hspace{-1cm}
\includegraphics[scale=0.5]{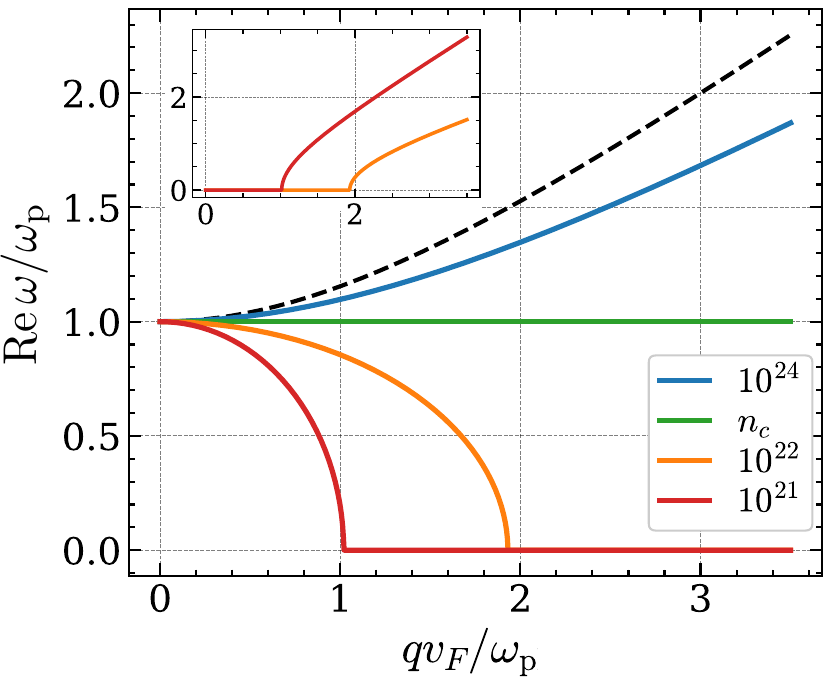}
\caption{(color online) Real part of the plasmon dispersion for different densities indicated in the caption in units of  $\text{cm}^{-3}$. The dashed black line corresponds to the result in the absence of exchange, in which case the normalized dispersion becomes density independent. For $n=n_c \sim 5\times 10^{22} \, \text{cm}^{-3}$ we have $\omega(q) = \omega_\text{p}$. The inset shows the corresponding imaginary parts.}
\label{plasmon_disp}
\end{figure}

 
In Fig.~\ref{plasmon_disp} we depict the plasmon dispersion relation assuming an electron gas with varying density. For large densities ($n\gtrsim 10^{28}\,\text{cm}^{-3}$) exchange effects become negligible as $\tilde{v}/v_F$ approaches zero. In this case, all curves fall within the same universal curve when plotted in dimensionless variables (dashed curve of Fig.~\ref{plasmon_disp}). For moderate densities, a depletion of plasmon frequency for finite momentum is noted. Moreover, there is a critical value of density $n_c \simeq 5\times 10^{22} \, \text{cm}^{-3}$ for which classical and quantum effects cancel each other and the dispersion becomes flat, $\omega(q) = \omega_\text{p}$. For densities above $n_c$ the dispersion is dominated by kinetic energy, which results in a monotonically increasing function of $q$. Below $n_c$ exchange effects dominate and the dispersion attains a maximum at $q=0$, becoming zero at $q_\text{max} =\omega_\text{p}/\sqrt{\tilde{v}^2/2-v_F^2/3}$ and imaginary afterwards. The purely imaginary part of the spectrum can be associated to static density patterns that become unstable, typical of dense plasmas.

\section{Applications}
In this section we return to the fluid model derived in Sec.~\ref{SecIV} and apply it to two different physical situations of interest. Our main goal is to highlight the consequences of the semiclassical corrections therein to the dynamics of light-matter systems, as well as to clarify under which conditions are these corrections important. To do that, we first study the effect of electron exchange on the growth rates of collective modes propagating in a homogeneous photon gas in contact with a solid-state plasma. In particular, we demonstrate that qualitative differences are found when the exchange potential is included. Then, we consider the case of an intense light beam propagating diffusely inside a dense astrophysical environment. We show that, for sufficiently large field intensities, photon and electron density perturbations become highly coupled, and, depending on the direction of propagation, these may give rise to slow-light modes or avoided crossing. We also discuss how these modes may be associated with photon-bubble turbulence.   

\subsection{Thermal light in degenerate plasmas: the plasmon-intensiton polariton}

Let us assume a homogeneous photon gas in equilibrium at temperature $T_\text{ph}$ in contact with a solid-state plasma with density  $n_0$. We use $\mathcal N(\mathbf r,\mathbf k,t) = \mathcal N_0(T_\text{ph}) + \delta \mathcal N(\mathbf r,\mathbf k,t)$ and $\mathcal W_e(\mathbf r,\mathbf k,t) = \mathcal W_{0}(\mathbf k) + \delta \mathcal  W_e(\mathbf r,\mathbf k,t)$, with $\mathcal N_0(\mathbf k)$ the Bose-Einstein function and $\mathcal W_{0}(\mathbf k)$ the Fermi-Dirac distribution of Eq.~\eqref{FDlimits} in  the limit $X \gg 1$. \par
\begin{figure}[t!]
\includegraphics[scale=0.5]{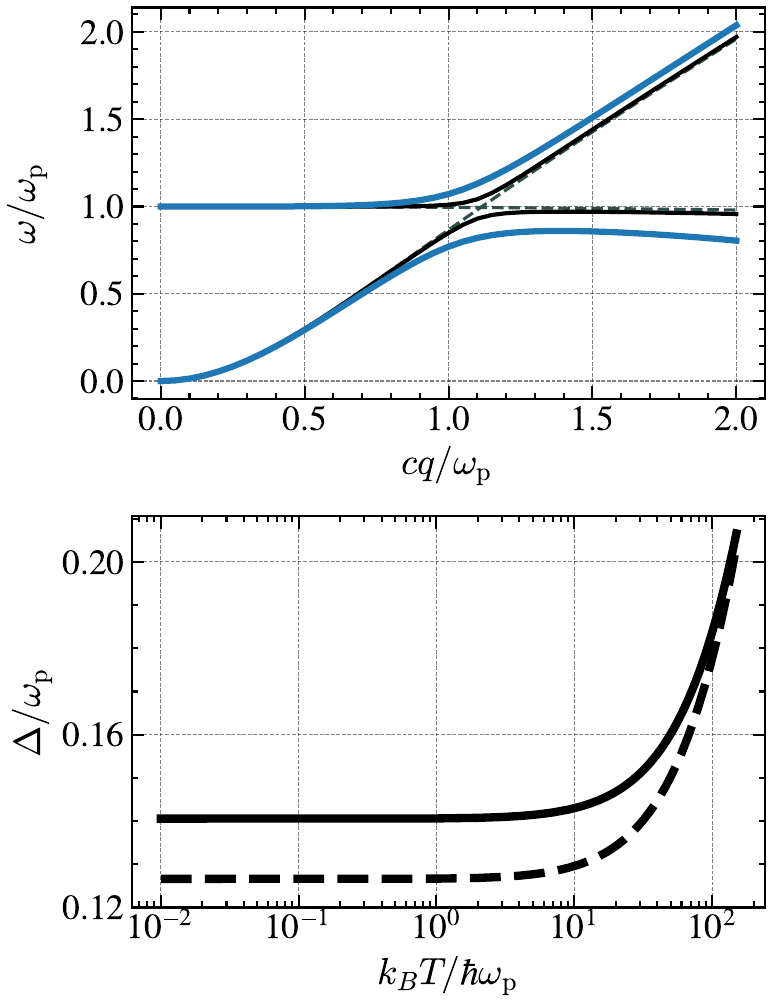}
\caption{(color online) Top panel: Dispersion relation of the polariton modes in Eq. \eqref{eq_dispA} resulting from the hybridization of the plasmon and photon-intensity modes. The dashed lines depict the bare modes for $v_{\rm eff}^2=-0.1 c^2$. The solid lines are obtained for $c_*=0.3 c$ (dark) and $c_*=0.5 c$ (light). Bottom panel: Rabi frequency for $v_{\rm eff}^2=-0.1 c^2$ (solid) and $v_{\rm eff}^2=0.1 c^2$ (dashed), with $c_\ast = 0.3c$. }
\label{fig_dispA}
\end{figure}

 
If the deviations from equilibrium are small, the response of the system can be separated into transverse and longitudinal contributions. Since the equilibrium electromagnetic fields are zero, then the transverse response is associated with electromagnetic waves propagating in the plasma with the usual dispersion. On the other hand, the equilibrium electromagnetic intensity $I_0$ is finite and is a function of $T_\text{ph}$ through Planck's distribution. Thus, after Fourier transforming Eqs.~\eqref{dtDN}, \eqref{dtDJ}, \eqref{dtDNp} and \eqref{dtDJp} in Fourier space, the longitudinal dielectric function can be extracted,
\begin{align}
        \varepsilon(q,\omega) &= 1 - \frac{\omega_\text{p}^2}{\omega^2-v_\text{eff}^2q^2} \nonumber 
    - \frac{c_p^4 q^4}{(\omega^2-v_\text{eff}^2q^2)(\omega^2-v_p^2q^2)},
    \label{eq_dispA}
\end{align}
with 
\begin{align}
   c_p(q) &= \left(\frac{1}{6\pi^2} \frac{\hbar \omega_\text{p}^5}{m_e n_0 c} \frac{\mathcal M_1 -\frac{1}{4}\mathcal M_2}{\sqrt{\omega_\text{p}^2 + c^2 q^2}}\right)^{1/4}, \quad \text{and}\nonumber \\
   v_p(q) &= \frac{c^2 q}{\sqrt{\omega_\text{p}^2 + c^2 q^2}}\sqrt{1+\frac{\omega_\text{p}^2}{\omega_\text{p}^2 + c^2 q^2}} \ .
\end{align}
Above, the coefficients $\mathcal M_\ell \equiv \mathcal M_\ell(\hbar\omega_\text{p}/k_B T_\text{p})$ are related to the Bose-Einstein distribution,
\begin{equation}
    \mathcal M_\ell(x) = \int_{0}^\infty dk \  \frac{k^{2\ell}}{(k^2+1)^\ell} \frac{1}{\text{exp}(x\sqrt{1+k^2})-1},
\end{equation}
and satisfy the relation $\mathcal{M}_{\ell +1}\leq \mathcal{M}_\ell$, assuring that $c_p$ is a positive quantity. The dispersion relation in Eq. \eqref{eq_dispA} displays two polariton modes, $\omega_\pm$, as depicted in Fig. \ref{fig_dispA},
\begin{equation}
\omega^2_\pm =\frac{1}{2} \left[ \omega_e^2+\omega_{\rm ph}^2 \pm\sqrt{\left(\omega_e^2-\omega_{\rm ph}^2\right)^2 +4c_p^4q^4}\right],  
\end{equation}
resulting from the hybridization of the electron density (plasmon) and light-intensity (intensiton) modes, with dispersions $\omega_e=\sqrt{\omega_p^2+v_{\rm eff}^2 q^2}$ and $\omega_{\rm ph}=v_p(q) q$, respectively. Notice that the intensiton dispersion discussed here, corresponding to fluctuations on the light intensity $I\sim \vert {\bf E}\vert^2\sim n_{\rm ph}$, differs from the more common electromagnetic (EM) mode $\sim {\bf E}$ in plasmas, $\omega=\sqrt{\omega_p^2+c^2k^2}$, in which the photon acquires an effective mass $m_{\rm ph}=\hbar\omega_p/c^2$ \cite{Swanson2003}. Conversely, in the long wavelenght limit $q\ll \omega_p/c$, the former reduces to $\omega_{\rm ph}\simeq \hbar q^2/2 \tilde m_{\rm ph}$, with $\tilde m_{\rm ph}=\hbar \omega_p/(2\sqrt{2}c^2)$ being the intensiton effective mass. \par The Hopfield coefficients $u_q$ and $v_q$, measuring the fraction of plasmon (intensiton) in the upper, $+$ (lower, $-$), polariton mode and satisfying the condition $\vert u_q\vert^2+\vert v_q\vert^2=1$, are given by
\begin{align}
u_q=\frac{\omega_+\omega_e-\omega_-\omega_{\rm ph}}{(\omega_e+\omega_{\rm ph})\sqrt{(\omega_e-\omega_{\rm ph})^2+4c_p^2 q^2}}\\
v_q=\frac{\omega_+\omega_{\rm ph}-\omega_-\omega_e}{(\omega_e+\omega_{\rm ph})\sqrt{(\omega_e-\omega_{\rm ph})^2+4c_p^2 q^2}}.
\end{align}
The crossing between the two modes $\omega_e$ and $\omega_{\rm ph}$ occurs at the wavevector $q_*\simeq \omega_p/\sqrt{c^2-v_{\rm eff}^2}$ and the strength of the coupling (avoided-crossing) is quantified by the Rabi frequency $\Delta=\omega_+-\omega_-\vert_{q=q_*}$ which, at leading order in $v_{\rm eff}$ and $c_*\equiv c_p(q=q_*)$, is given by
\begin{align}
\Delta &\simeq \frac{\omega_p}{2} \left[ 2-\sqrt{3}-\frac{v_{\rm eff}^2}{c^2}\left(\frac{2\sqrt{3}-3}{3}\right) \right. \nonumber\\ &+ \left.\frac{v_{\rm eff}^2c_*^4}{c^6}\left(\frac{27+16\sqrt{3}}{18} \right)\right].
    \label{eq_Rabi}
\end{align}
As illustrated in bottom panel of Fig. \ref{fig_dispA}, for low temperature degenerate Fermi gases, i.e $k_BT\lesssim \hbar \omega_p$, the coupling is enhanced as $v_{\rm eff}^2$ becomes negative due to exchange. Conversely, in the high temperature limit $k_BT\gg  \hbar \omega_p$, degeneracy effects become less relevant. This result suggests that this polariton effects are more likely to be observed in solid-state plasmas, in agreement with the features of the diagram in Fig. \ref{R_n_T}. Similar polariton quasiparticles, resulting from the hybridization of plasmons and other fields, have also been found in the context of axions in plasmas \cite{Tercas2018}. 

\subsection{Light instabilities in diffusive plasmas: the photon bubble case}

Next we consider a dense plasma composed by two fermionic species with a large mass difference. This corresponds, for example, to an electron--ion plasma, where the mass ratio is $\sim 10^{-4}$. In this case, the interaction between photons and charged particles will happen mostly with the lighter species, which we assume to be electrons in what follows. Moreover, the heavier (e.g., ionic) fluid is treated as a static background. \par 
Suppose that a photon beam centered at wavevector $\mathbf k_0$ and frequency $\omega_0 \equiv  \omega_{\mathbf k_0}$ enters the plasma in equilibrium. For simplicity, we assume that $k_0$ is sufficiently large such that $\omega_0 \simeq c k_0$. To determine the allowed collective modes of the system, we assume that all fluid quantities oscillate as $\sim e^{i(\mathbf q\cdot \mathbf r -\omega t)}$, where $\mathbf q$ is the collective wavevector, $\omega$ its corresponding frequency. In order to establish equations of state for the remaining variables, an ansatz for the fluctuating distribution functions is required. In the spirit of the linearized theory, we write
\begin{align}
	\delta \mathcal N(\mathbf r,\mathbf k,t) = \mathcal D(\mathbf k - \mathbf q)\delta n_\text{ph} e^{i(\mathbf q\cdot \mathbf r -\omega t)} , \label{PWignerExp} \\
	\delta \mathcal W_e(\mathbf r,\mathbf k,t) = \mathcal D(\mathbf k - \mathbf q)\delta n_e e^{i(\mathbf q\cdot \mathbf r -\omega t)},
\end{align} 
where $\mathcal D(\mathbf k)$ is a function highly peaked at zero and normalized to the volume, $\sum_{\mathbf k} \mathcal  D(\mathbf k) = V$, and $\delta n_{e, {\rm ph}}$ denotes the amplitude of density fluctuations. This form ensures that $V^{-1} \sum_{\mathbf k} \delta \mathcal N(\mathbf r,\mathbf k) = \delta n_\text{ph}e^{i(\mathbf q\cdot \mathbf r -\omega t)}$, and similarly for electrons. Taking $n_e=n_0+\delta n_e$ and $n_{\rm ph}=N_0+\delta n_{\rm ph}$, and neglecting second-order terms in the fluctuations, Eqs. \eqref{dtDN}, \eqref{dtDJ}, \eqref{dtDNp} and \eqref{dtDJp} yield
\begin{align}
\delta \mathbf{J}_e^{(2)} &\simeq -\frac{\omega_\text{p}^2}{2} \Bigg( \frac{c^2 \mathbf q}{\omega_{\mathbf q}^3} \delta n_\text{ph} +  \frac{N_0}{n_0}\frac{c^2 \mathbf k_0}{\omega_0^3}\delta  n_{e}\Bigg)e^{i(\mathbf q\cdot \mathbf r -\omega t)}, \label{EqState6} \\       
	\delta \mathbf{F}_e &\simeq  - i\frac{\omega_\text{p}^2}{2n_0}  \frac{\hbar}{\omega_{\mathbf q}} {\bf q} \, \delta n_\text{ph} e^{i(\mathbf q\cdot \mathbf r -\omega t)}. \label{EqState5}
\end{align} \par 
In turn, the vector potential is given by
\begin{equation}
	\mathbf A(\mathbf r,t) = \mathbf A_0 e^{i(\mathbf k_0\cdot \mathbf r - \omega_0 t)} + \delta \mathbf A(\mathbf r,t).
\end{equation} 
The first term gives rise to the (transverse) electric and magnetic fields of the unperturbed beam, $\mathbf{E}_0 = i\omega_0 \mathbf A_0 e^{i(\mathbf k_0\cdot \mathbf r - \omega_0 t)}$ and $\mathbf{B}_0 = i\mathbf k_0 \times  \mathbf A_0 e^{i(\mathbf k_0\cdot \mathbf r - \omega_0 t)}$. Similarly, $\delta \mathbf{E} = -\partial_t \delta \mathbf A$ and $\delta \mathbf{B} = \bm{\nabla}\times \delta \mathbf A$ are the field perturbations. \par 
When the unperturbed densities are sufficiently high, light propagation within the plasma becomes predominantly diffusive. In this regime, the photon density changes much more rapidly than the photon current, allowing the time derivative of the current to be neglected. In the opposite (convective) limit, the propagation of modes on top of the photon gas becomes dynamically unstable (modulational instability) \cite{Mendonca2000, Shukla2004, Marklund2005, Marklund2006, Tsintsadze2007, Singh_2013}. In the diffusive limit, $\mathbf J_\text{ph}$ might be related with the light intensity $I$ from Eq.~\eqref{dtDJp} by setting $\partial_t \mathbf{J}_\text{ph}=0$ and using $n_\text{ph} = I/( \hbar \omega c)$, with $\omega$ the light frequency for  mode $\mathbf k$. Then Eq.~\eqref{dtDNp} reduces to a diffusion equation for the intensity, 
\begin{equation}
\frac{\partial}{\partial t} I - \bm{\nabla}\cdot(D \bm{\nabla}I)  = - G I + S.
\end{equation}
Above, 
\begin{equation}
    D = \left(1+\frac{\Omega_e^2}{\omega^2}\right)^2 \frac{c^2 \omega}{\mathbf{k} \cdot \bm{\nabla} \Omega_e^2} \ \mathbf{k} \otimes \mathbf{k}
\end{equation}
is the diffusive tensor, depending on space and time via the local plasma frequency $\Omega_e=\Omega_e[n_e({\bf r},t)]$, and 
\begin{equation}
    G = \omega\left(1+\frac{\Omega_e^2}{\omega^2}\right) \frac{\bm{\nabla}^2 \Omega_e^2}{\mathbf k \cdot \bm{\nabla} \Omega_e^2} 
\end{equation}
denotes the absorption coefficient, both stemming from Compton-scattering events. 
\par 
For the plasma variables, it is convenient to decompose the total electron current in its transverse and longitudinal components, $\mathbf{J}_e = \mathbf{J}_e^\perp + \mathbf{J}_e^\parallel$. The unperturbed electron current $\mathbf J_{e,0}$ is finite and displays a space and time dependence due to Lorentz acceleration by the unperturbed fields, governed by
\begin{align}
	\frac{\partial}{\partial t} \mathbf{J}_{e,0}^\perp  & =  -\frac{e}{m_e}\Big( n_0 \mathbf{E}_0  + \mathbf{J}_{e,0}^\parallel \times \mathbf{B}_0 \Big),\\
	\frac{\partial}{\partial t} \mathbf{J}_{e,0}^\parallel  & =  -\frac{e}{m_e} \mathbf{J}_{e,0}^\perp \times \mathbf{B}_0 . 
 \end{align}
Because both $\mathbf{E}_0$ and $\mathbf{B}_0 $ contain $e^{-i\omega_0t}$, the solutions will be of the form $\mathbf{J}_{e,0}^\perp,\mathbf{J}_{e,0}^\parallel \sim e^{-i\omega_0t}$ representing fast electron drifts. Similarly, the evolution of fluctuating currents contains fast and slow contributions, with the fast scales being associated to the unperturbed beam. The latter average out for sufficiently large separation of scales, and we arrive at the linearized equations for the slow components:
\begin{align}
 & \frac{\partial}{\partial t} \delta \mathbf{J}_{e}^\perp  =  -\frac{en_0}{m_e}\delta \mathbf{ E}, \label{perpJ}\\
&	\frac{\partial}{\partial t} \delta \mathbf{J}_{e}^\parallel +\frac{1}{m_e}\bm \nabla \big(\delta P_e + \delta P^\text{F}_e\big)  \nonumber\\
& =\frac{n_0}{m_e}\Big( e \bm\nabla^2\delta \Phi^\text{H} +\delta \bm{f}_e^\text{F} +  \delta \mathbf{F}_e  \Big) .
\end{align}

\begin{figure}[t!]
\centering \hspace{-1.2cm}
\includegraphics[scale=0.6]{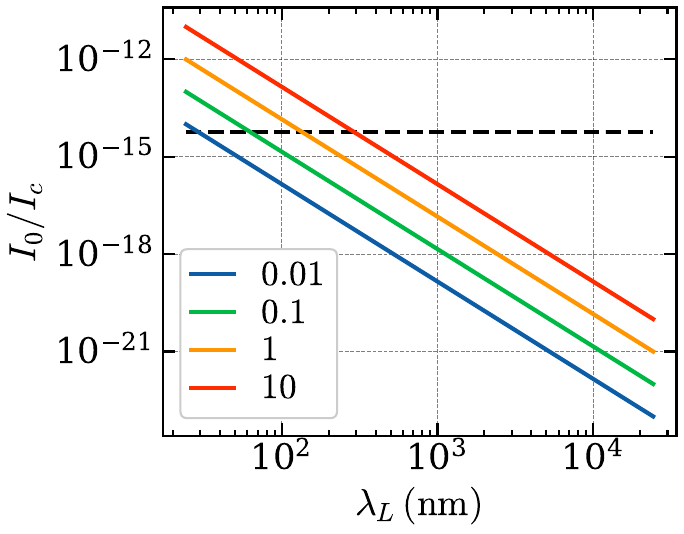}
\caption{(color online) Relation between the beam's normalized intensity and wavelength for fixed $\gamma$ (indicated in the caption). The dashed black line corresponds to the intensity above which electrons become relativistic, hence the area below line defines the region of validity of the model.}
\label{I0_lambda}
\end{figure}

After joining all the results, we are led to
\begin{equation}
	\begin{pmatrix}
		D_\perp(\mathbf q,\omega) & 0 \\
		0 & D_\parallel(\mathbf q,\omega)
	\end{pmatrix} 
	\begin{pmatrix}
		\delta\mathbf{U}_\perp(\mathbf q,\omega)\\
		\delta \mathbf{U}_\parallel(\mathbf q,\omega)
	\end{pmatrix}  = 0, \label{dispMatrix}
\end{equation}
where $\delta \mathbf{U}_\perp = (\delta \mathbf J_e^{\perp},\delta \mathbf  A)$ and $\delta \mathbf{U}_\parallel = (\delta n_e,\delta \mathbf J_e^{\parallel},\delta I)$ denote transverse and longitudinal quantities respectively, and $D_{\perp,\parallel}$ are matrices. Two of the roots are related to the transverse variables, and correspond to usual electromagnetic modes in plasmas, with dispersion $\omega = \pm \sqrt{\omega_\text{p}^2 + c^2 q^2}$. The three remaining modes are hybridizations between plasma and the laser intensity modes, similarly to the previous section for thermal light. These modes result from the secular equation $\text{det}D_\parallel(\mathbf q,\omega)=0$, which we can recast as
\begin{equation}
	\Big(\omega^2- \omega^2_{e}\Big) \Big(\omega-\widetilde{\omega}_\text{ph}\Big)+ \Gamma  = 0. \label{dispLongi}
\end{equation}
Here, $\widetilde{\omega}_\text{ph} = c^2 q^2 \left(1 - \frac{1}{2} \frac{ \omega_\text{p}^2}{\omega_{q}^2}\right)/\omega_{q}$ is the laser counterpart of the intensiton dispersion defined above for the thermal photon gas, and  
\begin{align}
	\Gamma( q,\omega) &= \gamma \cos\theta \Bigg(  \frac{\hbar \omega_\text{p}^2}{2 m_e c^2} \frac{cq}{\omega} - \frac{\omega_\text{p}^3}{\omega_{ q}^2} \, \Bigg)\frac{c^2 q^2 \omega }{\omega_q} 
\end{align}
denotes the coupling function. Moreover, $\theta$ is the angle between $\mathbf q$ and $\mathbf k_0$ and $\gamma$ is the laser-plasma coupling strength,
\begin{align}
	\gamma &= \frac{I_0}{I_c}\Bigg(\frac{\lambda_L}{\lambda_c} \Bigg)^3,
\end{align} 
with $I_{0} = c\hbar \omega_0  N_0$ being the unperturbed laser intensity, $\lambda_L = 2\pi/k_0$ the wavelength, $I_c = \epsilon_0 m_e^4 c^7/(2\hbar^2 e^2)\simeq 2.3 \times 10^{29}\, \text{W}/\text{cm}^2$ the Schwinger intensity and $\lambda_c = 2\pi \hbar /(m_ec)\simeq 2.4 \times 10^{-12} \, \text{m}$ the electron Compton wavelength. In Fig.~\ref{I0_lambda} the relation between $I_0$ and $\lambda_L$ is shown for several coupling strengths.\par 

\begin{figure}[t!]
\centering
\hspace{-0.1cm}
\includegraphics[scale=0.38]{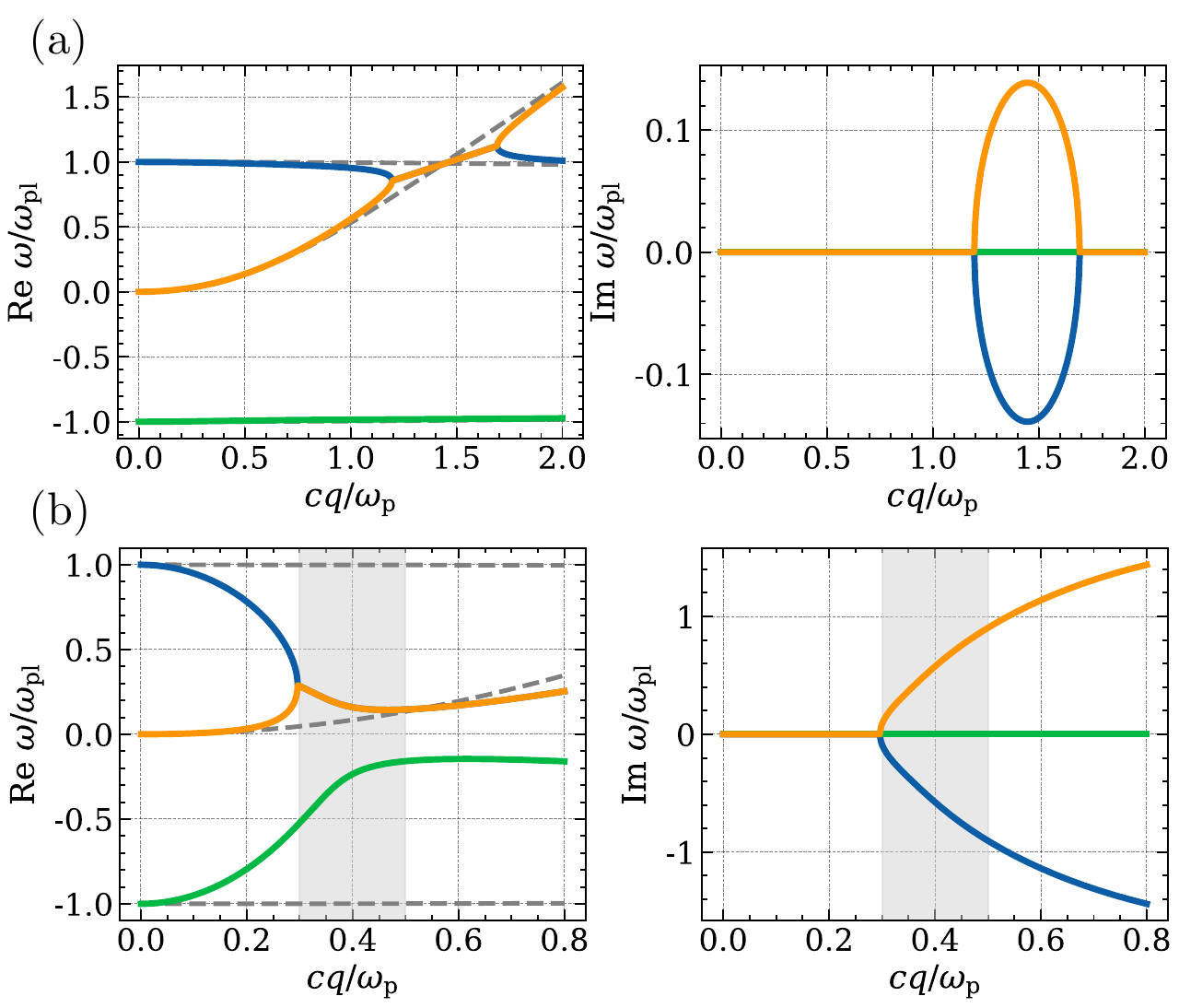}
\caption{(color online) Real and imaginary parts of hybridized light-matter modes with $n_0 = 10^{27} \, \text{cm}^{-3}$, $\theta = \pi$, and (a) moderate coupling $\gamma = 0.1$ or (b) strong coupling $\gamma =10$. At this densities degeneracy requires $T< 5 \times 10^7\,$K which is typically verified for stellar cores. Grey lines correspond to uncoupled modes $\pm \omega_e$ and $\widetilde{\omega}_\text{ph}$. The latter is associated with photon-density fluctuations, verifying $\sim  c^2 q^2 /(2\omega_\text{p})$ close to the origin and $\sim c q$ for large $q$. The nonzero curvature at $q=0$ is equivalent to a mass $m_p = \hbar \omega_\text{p}/c^2$. In panel (b) shaded grey areas mark regions of negative and small group velocities.}
\label{disp_fig1}
\end{figure}

In the limit $\gamma\to 0$, the plasmon and intensiton modes decouple. In this case the modes reduce to $\omega_{e}$ and $\widetilde{\omega}_\text{ph}$, depicted with grey lines in Fig.~\eqref{disp_fig1}. Uncoupled modes are attained for either small $I_0 \lambda_L^3$ or transverse propagation with respect to the unperturbed beam. Physically, a small value of $I_0 \lambda_L^3$ corresponds to the plasma becoming transparent to the beam ($\lambda_L \simeq 0$) or to a small number of photons ($I_0 \simeq 0$) such that nonlinear interactions become irrelevant. For finite values of $\gamma$, we observe two types of hybridizations between plasmons and intensitons, depending on the direction of propagation.

For modes propagating with nonzero parallel component with respect to the initial beam ($\cos\theta >1$), the coupling results in an avoided crossing similar as the one depicted in Fig. \eqref{fig_dispA}, with the strength of the coupling governed by $\gamma$. Conversely, for counter-propagating directions ($\cos\theta <1$), the plasmon and the intensiton modes coalesce and become unstable [panels (b) and (c) of Fig.~\ref{disp_fig1}]. This instability is of electromagnetic nature and lifts the degeneracy instability present for zero coupling, by modifying the plasmon branch in the long-wavelength region. For sufficiently small, but nonzero, coupling [panel (a)], the coupled modes overlap close to the plasma frequency and the unstable part of the spectrum is limited. This phenomenon is reminiscent to the photon bubbling instability. Photon bubbles are light-density inhomogeneities that are produced in radiation-dominated (or optically dense) media, such as accretion disks and dust clouds surrounding high-mass stars \cite{Begelman2006, Turner2007}, and have recently been observed in ultracold matter \cite{Mendonca2012, Giampaoli2021}. Conversely, if the coupling is strong [panel (b)], one of the modes attains a maximum, whose neighborhood defines a region of zero group velocity and positive imaginary part. Close to the maximum, the latter is essentially an intensiton mode, and can thus be associated with slow-light propagation across the dense plasma due to multiple scattering. 

\section{Conclusion}

In this work, we developed a comprehensive quantum kinetic framework to describe interactions between light and matter in degenerate systems. By treating both photons and matter particles on equal footing, we incorporated both linear and nonlinear light-matter couplings, retaining quantum corrections up to second order in coupling constants. This approach goes beyond the semiclassical treatments often discussed in the literature by including nonlinearities from quantum fluctuations and exchange effects, with the latter represented by a modified Fock potential, Eq.~\eqref{closureFock}. Our kinetic equations capture the key quantum features of light-matter interactions through Eqs. \eqref{lambda1}$-$\eqref{C2_sin}, extending the framework to account for photon absorption, emission, and Compton scattering across a wide range of parameters. We also show how the classical Lorentz dynamics can be recovered from the underlying quantum-field theory, reading as the classical (i.e., $\hbar$-independent) contribution of quantum absorption and emission processes. 

Our results demonstrate that the Fock potential introduces significant corrections in the long-wavelength limit of the plasmon dispersion relation, particularly for high densities. These results address gaps in previous studies, where exchange effects were modeled using an approximate classical potential, showing that this treatment is incomplete as it neglects additional renormalizations of single-particle velocity as derived here. The latter introduce quantum contributions to the macroscopic currents and pressures, which modify the plasmon spectrum leading to new dispersion branches that could be observed experimentally in systems with high electron degeneracy. These can be, e.g., solid-state plasmas or warm-dense states of matter such as astrophysical objects. We found that plasmons hybridize with the photon-intensity (intensiton) mode of a thermal photon gas, forming polaritonic quasiparticles, where coupling is influenced by exchange effects. Additionally, considering a light beam propagating in a dense plasma under diffusive conditions, we predicted the appearance of photon bubbles. 

Our results can be extended to various scenarios where high-field effects and quantum degeneracy coexist. A potential application is laser-plasma interactions, where strong field gradients and degenerate electron populations could enhance harmonic generation and trigger novel plasma instabilities \cite{PhysRevA.82.043414,Gorlach}. Moreover, our model is also capable of addressing photon dynamics in highly magnetized environments, where photon-photon interactions mediated by the plasma could lead to magneto-optical structures and complex field configurations \cite{10.1063/1.4913435}. Future research could involve extending the present kinetic theory to spinor fields to account for spin-plasma interactions \cite{PhysRevD.101.076022}, or investigating kinetic effects in high-energy astrophysical scenarios. Furthermore, the quantum higher-order terms in the ponderomotive forces emerging from our theory may impact photon condensation scenarios, as quantum corrections suppress the classical Zel\^{a}ovich instability threshold, altering phase transition dynamics. These findings suggest that our results could be used to predict the onset of photon condensation in systems not well-described by classical or semiclassical models, such as those involving quantum matter or photon nonlinearities \cite{PhysRevLett.111.100404,Barland:21,PhysRevB.76.201305}. Also, it will be instructive to explore the coupling between light-intensity and electron-density modes in photon BECs, in which the condensate mode features an acoustic (Bogoliubov) mode $\omega\sim q$ associated to the spontaneous breaking of the U(1) symmetry \cite{plasmaBEC1}. \par

In conclusion, our approach not only provides a more accurate description of photon kinetics in degenerate Fermi systems but also opens new avenues for exploring the complex dynamics of light-matter interactions under extreme conditions. The results offer a strong foundation for future studies of quantum plasma behavior and their potential applications in both fundamental physics and practical technologies.

\section{Acknowledgments}
J. L. F. and H. T. acknowledge Funda\c{c}\~{a}o para a Ci\^{e}ncia e a Tecnologia (FCT-Portugal) through the Grants No. PD/BD/135211/2017, UI/BD/151557/2021, and through Contract No. CEECIND/00401/2018 and Project No. PTDC/FIS-OUT/3882/2020, respectively.

\appendix

\begin{widetext}

\section{Derivation of Wigner equations}\label{appW} 
In this Appendix we outline the derivation of Eqs.~\eqref{photon1} and \eqref{plasmaWigner1} of the main text. The important correlators are defined as $A_{\mathbf k,\mu} \equiv \langle \hat a_{\mathbf k,\mu}\rangle$, $N_{\mathbf k,\mu,\mathbf k',\mu'} \equiv \langle \hat a^\dagger_{\mathbf k,\mu} \hat a_{\mathbf k',\mu'} \rangle $ and $W_{\alpha, \mathbf k,\mathbf k'} \equiv \langle \hat c^\dagger_{\alpha,\mathbf k} \hat c_{\alpha,\mathbf k'} \rangle$, which have a one-to-one correspondence with kinetic functions through inverse Fourier transformations:
\begin{align}
	A_{\mathbf k,\mu}(t) &= C_{\mathbf k}\bm{\varepsilon}_{\mathbf k,\mu}\cdot\frac{1}{V}  \int d\mathbf r \  e^{-i\mathbf r\cdot \mathbf k} \ \mathbf{A}^{(+)}(\mathbf r,t), \label{B_Expanded}\\
	N_{\mathbf k,\mu,\mathbf k',\mu'}(t) &= \frac{1}{V} \int d\mathbf r \  e^{i\mathbf r\cdot (\mathbf k-\mathbf k')} \mathcal N_{\mu,\mu'}\left(\mathbf r,\frac{\mathbf k+\mathbf k'}{2},t\right),\label{n_Expanded}\\
	W_{\alpha,\mathbf k,\mathbf k'}(t) &= \frac{1}{V} \int d\mathbf r  \ e^{i\mathbf r\cdot(\mathbf k-\mathbf k')}\mathcal W_\alpha\left(\mathbf r,\frac{\mathbf k+\mathbf k'}{2},t \right). \label{W_Expanded}
\end{align}
Above, we defined $C_{\mathbf k} = (2\epsilon_0\omega_{\mathbf k} V/\hbar)^{1/2}$. \par
In order to establish equations of motion for distribution functions, it is necessary to derive the equations for the corresponding correlators. By doing so, higher-order correlators appear due to nonzero quantum fluctuations, as a result of interactions. These fluctuations are defined as
\begin{align}
	\delta \hat a_{\mathbf k,\mu} &\equiv  \hat a_{\mathbf k,\mu} - A_{\mathbf k,\mu},\label{photonFluc}\\
	\delta \hat c^\dagger_{\alpha,\mathbf k} \hat c_{\alpha,\mathbf k'} &\equiv \hat c^\dagger_{\alpha,\mathbf k} \hat c_{\alpha,\mathbf k'} - W_{\alpha,\mathbf k,\mathbf k'}, \label{fermFluc}
\end{align}
and verify $\langle \delta \hat a_{\mathbf k,\mu} \rangle = \langle \delta \hat c^\dagger_{\alpha,\mathbf k} \hat c_{\alpha,\mathbf k'} \rangle = 0$. Note that $\delta \hat c^\dagger_{\alpha,\mathbf k} = \hat c^\dagger_{\alpha,\mathbf k}$ since $\langle \hat c^\dagger_{\alpha,\mathbf k}\rangle =0$, so Eq.~\eqref{fermFluc} defines the first nonvanishing fermionic fluctuation. Upon neglecting all fluctuations, we are led to the classical Hamiltonian, where each operator is replaced by its expectation value. The classical Hamiltonian can then be used to determined all classical electrodynamical laws. \par 
One can show that quantum fluctuations contribute with higher-order corrections in the coupling constants. Therefore, the classical limit applies if interactions are sufficiently weak when compared to the kinetic-energy scales. In this work, we are interested in situations where the fluctuations of the photon field are important, which correspond to cases where the light intensity is strong enough so that the nonlinear Hamiltonian of Eq.~\eqref{Hint2} needs to be taken into account. In such cases, neglecting the photon fluctuations to all orders leads to an incomplete description because taking $\hat a_{\mathbf k,\mu} = A_{\mathbf k,\mu}$ in Eq.~\eqref{Hint1} requires that one neglects Eq.~\eqref{Hint2} to maintain consistency. In other words, taking $\delta \hat a_{\mathbf k,\mu}=0$ in Eq.~\eqref{Hint1} introduces an error of order $M^2$, while doing the same in Eq.~\eqref{Hint2} is valid to order $D$. However, since $M$ depends on fermionic degrees of freedom as well, there will be scattering events at high intensities for which $M^2/D \sim 1$, thus rendering photon fluctuations important. This inconsistency can be solved by introducing the photon Wigner function. On the other hand, fermionic fluctuations can be neglected because Coulomb correlations are typically unimportant in plasmas, even in the quantum regime. \par 
With this in mind, we take $\langle \hat a_{\mathbf q,\mu} \hat c_{\alpha,\mathbf k}^\dagger \hat c_{\beta,\mathbf k'} \rangle = \delta_{\alpha,\beta}A_{\mathbf q,\mu} W_{\alpha,\mathbf k,\mathbf k'}$, $\langle \hat a_{\mathbf q',\mu'}^\dagger \hat a_{\mathbf q,\mu} \hat c_{\alpha,\mathbf k}^\dagger  \hat c_{\beta,\mathbf k'} \rangle = \delta_{\alpha,\beta}N_{\mathbf q',\mu',\mathbf q,\mu} W_{\alpha,\mathbf k,\mathbf k'}$ and $\langle \hat c_{\alpha,\mathbf k}^\dagger \hat c_{\beta,\mathbf k'}^\dagger \hat c_{\beta,\mathbf q'}\hat c_{\alpha,\mathbf q}\rangle = W_{\alpha,\mathbf k,\mathbf q}W_{\beta,\mathbf k',\mathbf q'} - \delta_{\alpha,\beta}W_{\alpha,\mathbf k,\mathbf q'}W_{\beta,\mathbf k',\mathbf q}$, which results in a closed system of equations for the correlators:
\begin{align}
&\frac{\partial}{\partial t} N_{\mathbf k,\mu,\mathbf k',\mu'} = \frac{i}{\hbar} \sum_{\mathbf q}\big( \mathcal G_{\mathbf k,\mathbf q} N_{\mathbf q,\mu,\mathbf k',\mu'} - N_{\mathbf k,\mu,\mathbf q,\mu'} \mathcal G_{\mathbf q,\mathbf k'} \big) +\frac{i}{\hbar} \sum_{\mathbf q,\alpha} \Big( M_{\mathbf k,\mathbf q}^{\mu,\alpha} A_{\mathbf k',\mu'}  W_{\alpha,\mathbf k + \mathbf q,\mathbf q} - M_{\mathbf k',\mathbf q}^{\mu',\alpha} A_{\mathbf k,\mu}^\ast W_{\alpha,\mathbf k'+\mathbf q,\mathbf q}^\ast\Big)  \nonumber \\
& \quad \quad + \frac{i}{\hbar} \sum_{\mathbf q,\mathbf q',\alpha,\sigma} \Big( D_{\mathbf k,\mathbf  q}^{\mu,\sigma,\alpha} N_{\mathbf q,\sigma,\mathbf k',\mu'} W_{\alpha,\mathbf k+\mathbf q',\mathbf q+\mathbf q'}  - D_{\mathbf k',\mathbf  q}^{\mu',\sigma,\alpha}  N_{\mathbf k,\mu,\mathbf q,\sigma}W_{\alpha, \mathbf q+\mathbf q',\mathbf k'+\mathbf q'} \Big), \label{dtNN}\\
&\frac{\partial}{\partial t} W_{\alpha,\mathbf k,\mathbf k'} = \frac{i}{\hbar} \sum_{\mathbf q}\big( \mathcal E_{\alpha,\mathbf k,\mathbf q} W_{\alpha,\mathbf q,\mathbf k'} - W_{\alpha,\mathbf k,\mathbf q}\mathcal E_{\alpha,\mathbf q,\mathbf k'} \big) + \frac{i}{\hbar} \sum_{\mathbf q,\mu } \Big(M_{\mathbf q,\mathbf k}^{\mu,\alpha} \big[ A_{\mathbf q,\mu}W_{\alpha,\mathbf k + \mathbf q, \mathbf k'} + A_{\mathbf q,\mu}^\ast W_{\alpha,\mathbf k', \mathbf k - \mathbf q}^\ast \big] \nonumber \\
&\quad \quad - M_{\mathbf q, \mathbf k'}^{\mu,\alpha} \big[ A_{\mathbf q,\mu} W_{\alpha,\mathbf k, \mathbf k' - \mathbf q} + A_{\mathbf q,\mu}^\ast W_{\alpha, \mathbf k' +\mathbf q, \mathbf k}^\ast \big] \Big)
	 + \frac{i}{\hbar} \sum_{\mathbf q,\mathbf q',\mu,\mu'} D_{\mathbf q,\mathbf q'}^{\mu,\mu',\alpha}N_{\mathbf q,\mu,\mathbf q',\mu'}\Big( W_{\alpha,\mathbf k,\mathbf k'+\mathbf q-\mathbf q'}- W_{\alpha,\mathbf k-\mathbf q+\mathbf q',\mathbf k'} \Big).\label{dtFF}
\end{align}
Above, 
\begin{equation}
	\mathcal G_{\mathbf k,\mathbf k'} = \hbar\omega_{\mathbf k}\delta_{\mathbf k,\mathbf k'} 
\end{equation}
and 
\begin{equation}
	\mathcal E_{\alpha, \mathbf k,\mathbf k'} = \xi_{\alpha,\mathbf k}\delta_{\mathbf k,\mathbf k'} + Q_\alpha \Phi^\text{H}_{\mathbf k' - \mathbf k} + \Phi^\text{F}_{\alpha, \mathbf k,\mathbf k'}
\end{equation}
are, respectively, the bosonic and fermionic single-particle Hamiltonians, while
\begin{equation}
	\Phi^\text{H}_{\mathbf k} = \sum_{\alpha} \frac{Q_\alpha}{\epsilon_0|\mathbf k|^2} n_{\alpha,\mathbf k}
\end{equation}
corresponds to the Hartree potential in Fourier space, with $n_{\alpha,\mathbf k}\equiv \int d\mathbf r \ e^{-i\mathbf k \cdot \mathbf r} \langle \hat   \Psi^\dagger_\alpha(\mathbf r) \hat \Psi_\alpha(\mathbf r)\rangle = \sum_{\mathbf k'} W_{\alpha,\mathbf k+\mathbf k',\mathbf k'}$ being the Fourier transform of the density expectation value. Additionally, 
\begin{equation}
	 \Phi^\text{F}_{\alpha, \mathbf k,\mathbf k'} = -\sum_{\mathbf q} U^{\alpha,\alpha}_{\mathbf q} W_{\alpha,\mathbf k + \mathbf q,\mathbf k' + \mathbf q} 
\end{equation}
denotes the Fock contribution to the Coulomb interaction, having no classical analogue due to the off-diagonal terms $\mathbf k \neq \mathbf k'$. For this reason, $\Phi^\text{F}$ provides phase-space corrections not only to the electrostatic potential but also to the single-particle velocity [c.f. Eq.~\eqref{kinPlasma}]. \par
From Eqs.~\eqref{dtNN} and \eqref{dtFF} a closed kinetic model can be derived, leading to Eqs.~\eqref{photon1} and \eqref{plasmaWigner1} of the main text. The latter contain a term describing single-particle dynamics, which we denote by $\mathcal K$, plus light-matter collisions related to the $M(D)$ interaction matrix elements. The latter are denoted by $\Lambda^{1(2)}$ for bosons and $\mathcal C^{1(2)}$ for fermions, and are calculated in the next section.

\subsection{Photon dynamics}
After replacing the single-particle terms of Eq.~\eqref{dtNN} into the time derivative of Eq.~\eqref{n_Expanded} and doing simple substitutions, we find the single-particle contribution
\begin{align}
\mathcal K[\mathcal N_{\mu,\mu'}, \mathcal G]  &= \frac{i}{\hbar}\frac{1}{V^2}\sum_{\mathbf q,\mathbf q'}\int d\mathbf r_1\int d\mathbf r_2 \ e^{i\mathbf q\cdot (\mathbf r-\frac{\mathbf r_1}{2}-\frac{\mathbf r_2}{2})} e^{i\mathbf k\cdot (\mathbf r_2 - \mathbf r_1)}e^{i\mathbf q'\cdot (\mathbf r_1 - \mathbf r_2)} \Big[\mathcal N_{\mu,\mu'}(\mathbf r_1,\mathbf k_+)\mathcal G(\mathbf r_2, \mathbf k_-)\nonumber \\
	& - \mathcal G(\mathbf r_1, \mathbf k_+)\mathcal N_{\mu,\mu'}(\mathbf r_2,\mathbf k_-)\Big],
\end{align}
where $\mathcal G(\mathbf r,\mathbf k) = \sum_{\mathbf q} e^{i\mathbf q\cdot \mathbf r} \mathcal G_{\mathbf k - \mathbf q/2,\mathbf k + \mathbf q/2}$ is a Wigner transform and $\mathbf k_{\pm} = \frac{\mathbf k}{2} + \frac{\mathbf q'}{2} \pm \frac{\mathbf q}{4}$. Next we introduce the change of variables $\mathbf k_1 = \mathbf q' - \mathbf k + \frac{\mathbf q}{2}$ and $\mathbf k_2 = \mathbf q' - \mathbf k - \frac{\mathbf q}{2}$ to get
\begin{align}
	\mathcal K[\mathcal N_{\mu,\mu'}, \mathcal G]  &= \frac{i}{\hbar}\frac{1}{V^2}\sum_{\mathbf k_1,\mathbf k_2}\int d\mathbf r_1\int d\mathbf r_2 \  e^{i\mathbf k_1\cdot(\mathbf r - \mathbf r_1)}  e^{i\mathbf k_2\cdot(\mathbf r_2 - \mathbf r)}\Big[\mathcal N_{\mu,\mu'}(\mathbf r_1,\mathbf k + \mathbf k_1/2)\mathcal G(\mathbf r_2, \mathbf k + \mathbf k_2/2) \nonumber \\
	& -\mathcal G(\mathbf r_1, \mathbf k + \mathbf k_1/2)\mathcal N_{\mu,\mu'}(\mathbf r', \mathbf k + \mathbf k_2/2)\Big].
\end{align} \par 
Our goal now is to transform the shifts in momentum coordinates into differential operators acting on the shifted functions. This is achieved with the help of the following identities, 
\begin{align*}
  f(\mathbf r + \mathbf s) &= \exp(\mathbf s\cdot \overrightarrow{\bm{\nabla}}_\mathbf{r})f(\mathbf r),\\
  \int d\mathbf r f(\mathbf r) \exp(\mathbf s\cdot \overrightarrow{\bm{\nabla}}_\mathbf{r})g(\mathbf r) &= \int d\mathbf r f(\mathbf r) \exp(-\mathbf s\cdot \overleftarrow{\bm{\nabla}}_\mathbf{r})g(\mathbf r),
\end{align*}
which can be used to identify two delta functions $\delta(\mathbf r) = V^{-1}\sum_{\mathbf k} e^{i\mathbf k\cdot \mathbf r}$. By performing the spatial integrations we arrive at the desired result 
\begin{align}
   \mathcal K[\mathcal N_{\mu,\mu'}, \mathcal G]  =  \mathcal N_{\mu,\mu'}(\mathbf r,\mathbf k) \frac{2}{\hbar}\sin\left(\frac{1}{2} \overleftarrow{\bm{\nabla}}_\mathbf{r} \cdot \overrightarrow{\bm{\nabla}}_\mathbf{k} - \frac{1}{2}  \overleftarrow{\bm{\nabla}}_\mathbf{k} \cdot \overrightarrow{\bm{\nabla}}_\mathbf{r}\right)\mathcal G(\mathbf r,\mathbf k). \label{mathcalK}
\end{align}
Since $\mathcal G_{\mathbf k,\mathbf k'}$ is diagonal, then $\mathcal G(\mathbf r,\mathbf k) = \hbar \omega_{\mathbf k}$ has no space dependence (i.e., there is no direct photon-photon potential) which simplifies the differential series above. The arrows in each differential operator refer to the direction of application. \par
Next, we focus on the linear light-matter contribution $\Lambda^{(1)}_{\mu,\mu'}$, which also reads as a functional of kinetic functions, although we do not write this dependence explicitly to ease the notation. It is defined as
\begin{equation}
	\Lambda^{(1)}_{\mu,\mu'} = \frac{i}{\hbar} \sum_{\mathbf q,\mathbf q',\alpha} e^{i\mathbf q\cdot \mathbf r} M^{\mu,\alpha}_{\mathbf k-\frac{\mathbf q}{2},\mathbf q'} A_{\mathbf k+\frac{\mathbf q}{2},\mu'} W_{\alpha,\mathbf k+\mathbf q' - \frac{\mathbf q}{2},\mathbf q'} + \text{h.c.},
\end{equation}
where 'h.c.' stands for conjugation with respect to the polarization indices. After introducing the kinetic functions, we may rewrite 
\begin{align}
\Lambda^{(1)}_{\mu,\mu'} &= 	\frac{i}{\hbar}\frac{1}{V^2}\sum_{\alpha,\mathbf q, \mathbf q'} \Big( -\frac{Q_\alpha}{m_\alpha}\Big) \int d\mathbf r_1 \int d\mathbf r_2 \ e^{i\mathbf q\cdot (\mathbf r -\frac{\mathbf r_1}{2}-\frac{\mathbf r_2}{2})} e^{i\mathbf k\cdot (\mathbf r_2 -\mathbf r_1)} \left(\frac{\omega_{\mathbf k + \mathbf q/2}}{\omega_{\mathbf k - \mathbf q/2}}\right)^{1/2}\nonumber \\
& \times \Big( \hbar \mathbf q' \, \Big| \, \bm{\varepsilon}_{\mathbf k - \mathbf q/2,\mu} \otimes \bm{\varepsilon}_{\mathbf k + \mathbf q/2,\mu'} \, \Big| \, \mathbf A^{(+)}(\mathbf r_1) \Big)  \mathcal W_\alpha(\mathbf r_2, \mathbf q' + \mathbf k/2 - \mathbf q/4) + \text{h.c.} \ ,
\end{align}
where we used the notation $\big(\bm a \big| \mathcal M \big| \bm  b\big)  \equiv \sum_{i,j} a_i \mathcal M_{i,j} b_j$, $\mathcal M$ being a square matrix with elements $\mathcal M_{i,j}$ and $\bm a,\bm b$ vectors. From the property $\bm{\varepsilon}_{\mathbf k,\mu} \cdot \mathbf k = 0$ and further algebraic manipulations we get 
\begin{align}
\Lambda^{(1)}_{\mu,\mu'} &= 	-\frac{i}{\hbar}\frac{1}{V}\sum_{\mathbf q} \int d\mathbf z \int d\mathbf s \ e^{i\mathbf q\cdot(\mathbf r-\mathbf z)+i\mathbf k\cdot \mathbf s}\left(\frac{\omega_{\mathbf k + \mathbf q/2}}{\omega_{\mathbf k - \mathbf q/2}}\right)^{1/2}\Big( \bm j(\mathbf z + \mathbf s/2) \, \Big| \, \bm{\varepsilon}_{\mathbf k - \mathbf q/2,\mu} \otimes \bm{\varepsilon}_{\mathbf k + \mathbf q/2,\mu'} \, \Big| \, \mathbf A^{(+)}(\mathbf z - \mathbf s/2) \Big)  + \text{h.c.}
\end{align}
where $\bm j(\mathbf r,t)$ is the total electric current of the plasma. Finally, the shifts with respect to $\mathbf q$ can be transformed into differential operators by making use of the previous identities. Then, by replacing $\mathbf A$ by the electric field, we arrive at
\begin{align}
\Lambda^{(1)}_{\mu,\mu'} &= 	-\frac{1}{\hbar} \int d\mathbf s \ e^{i\mathbf k\cdot \mathbf s}\Big( \bm j(\mathbf r + \mathbf s/2) \, \Big| \, \overleftrightarrow{\mathcal U}_{\mu,\mu'} \, \Big| \, \mathbf E^{(+)}(\mathbf r - \mathbf s/2) \Big)  + \text{h.c.}\ ,
\end{align}
where we defined the matrix operator $\overleftrightarrow{\mathcal U}_{\mu,\mu'}$ as 
\begin{equation}
	\overleftrightarrow{\mathcal U}_{\mu,\mu'} =  \frac{\bm{\varepsilon}_{\mathbf k,\mu} }{\sqrt{\omega_{\mathbf k}}}\exp\left(\frac{i}{2} \overleftrightarrow{\bm \nabla}_{\mathbf{r}}^{(+)} \cdot  \overleftrightarrow{\bm \nabla}_{\mathbf{k}}^{(-)}\right) \frac{\bm{\varepsilon}_{\mathbf k,\mu'} }{\sqrt{\omega_{\mathbf k}}} , \label{Uoperator}
\end{equation}
and $\overleftrightarrow{\bm \nabla}^{(\pm)}\equiv \overleftarrow{\bm \nabla} \pm \overrightarrow{\bm \nabla}$. Note that the spatial derivatives of $\overleftrightarrow{\mathcal U}_{\mu,\mu'}$ act on both $\bm j$ and $\mathbf E$, while the momentum derivatives act only inside $\overleftrightarrow{\mathcal U}_{\mu,\mu'}$, and not on the factor $ e^{i\mathbf k\cdot \mathbf s}$. In the main text we also defined the trace $\overleftrightarrow{\mathcal U}\equiv \sum_{\mu} \overleftrightarrow{\mathcal U}_{\mu,\mu}$.  \par
Similar manipulations can be used to arrive at 
\begin{align}
	\Lambda^{(2)}_{\mu,\mu'} &= \frac{i}{\hbar} \sum_{\alpha,\sigma} \int d\mathbf r_1 \widetilde{D}^{\mu,\sigma,\alpha}(\mathbf r-\mathbf r_1,\mathbf k)  \exp\left( \frac{i}{2} \overrightarrow{\bm{\nabla}}_{\mathbf r_1} \cdot \overleftrightarrow{\bm{\nabla}}_{\mathbf k}^{(+)} \right) \mathcal N_{\sigma,\mu'}(\mathbf r,\mathbf k) n_\alpha(\mathbf r_1) + \text{h.c.}, \label{Lambda2}
\end{align}
where $n_\alpha$ is the density of $\alpha$ particles and
\begin{equation}
	\widetilde{D}^{\mu,\mu',\alpha}(\mathbf r,\mathbf q) =  \sum_{\mathbf q'}e^{i \mathbf q' \mathbf r} D^{\mu,\mu',\alpha}_{\mathbf q - \mathbf q',\mathbf q} \label{FouriC}
\end{equation} 
is the Fourier transform of the scattering element. The spacial dependence of $\widetilde{D}^{\mu,\mu',\alpha}(\mathbf r,\mathbf q)$ traduces the nonlocality of light-matter scattering, i.e., it determines how a finite plasma density at position $\mathbf r_1$ can scatter light at some different position $\mathbf r$. A local scattering term is recovered when the wavelength of light is sufficiently larger than the plasma characteristic length, such that $\widetilde{D}(\mathbf r-\mathbf r_1)$ becomes proportional to $\delta(\mathbf r-\mathbf r_1)$ (see Sec.~\ref{HFlimit}).

\subsection{Matter dynamics}
Since the structure of the single-particle contribution of Eq.~\eqref{dtFF} is the same as that of Eq.~\eqref{dtNN}, then we may simply obtain one from the other with the appropriate substitutions, i.e., 
\begin{align}
   \mathcal K[\mathcal  W_\alpha, \mathcal E_\alpha]  =  \mathcal  W_\alpha(\mathbf r,\mathbf k) \frac{2}{\hbar}\sin\left(\frac{1}{2} \overleftarrow{\bm{\nabla}}_\mathbf{r} \cdot \overrightarrow{\bm{\nabla}}_\mathbf{k} - \frac{1}{2}  \overleftarrow{\bm{\nabla}}_\mathbf{k} \cdot \overrightarrow{\bm{\nabla}}_\mathbf{r}\right)\mathcal E_\alpha(\mathbf r,\mathbf k). \label{mathcalK2}
\end{align}
In this case, however, $\mathcal E_\alpha(\mathbf r,\mathbf k)$ depends on $\mathbf r$ as well, due to the Coulomb potential. \par 
The linear light-matter contribution is defined as
\begin{align}
\mathcal C^{(1)}_\alpha = \frac{i}{\hbar} \sum_{\mathbf q,\mathbf q',\mu} e^{i\mathbf q\cdot \mathbf r} A_{\mathbf q',\mu}^{\ast}\Big(M_{\mathbf q',\mathbf k - \frac{\mathbf q}{2}}^{\mu,\alpha}W_{\alpha,\mathbf k - \mathbf q' - \frac{\mathbf q}{2},\mathbf k + \frac{\mathbf q}{2}} - M_{\mathbf q',\mathbf k + \frac{\mathbf q}{2}}^{\mu,\alpha} W_{\alpha,\mathbf k - \frac{\mathbf q}{2},\mathbf k +  \mathbf q' + \frac{\mathbf q}{2}} \Big)  + \text{c.c}\ ,
\end{align}
where 'c.c.' denotes the complex conjugate. In terms of kinetic variables, this reads
\begin{align}
\mathcal C^{(1)}_\alpha &= \frac{Q_\alpha}{m_\alpha}\frac{i}{\hbar } \frac{1}{V^2}\sum_{\mathbf q,\mathbf q'} \int d\mathbf r_1 \int d\mathbf r_2 \ e^{i\mathbf q \cdot (\mathbf r - \mathbf r_2)}  e^{i\mathbf q'\cdot(\mathbf r_1 - \mathbf r_2)} \Big(\frac{\hbar\mathbf q}{2} -\hbar\mathbf k\Big) \cdot \mathbf A^{(-)}(\mathbf r_1) \Big[ \mathcal W_\alpha\Big(\mathbf r_2,\mathbf k - \frac{\mathbf q'}{2}\Big) \nonumber \\
&- \mathcal W_\alpha\Big(\mathbf r_2,\mathbf k + \frac{\mathbf q'}{2}\Big)\Big] + \text{c.c.} \  ,
\end{align}
where the explicit form of $M_{\mathbf k,\mathbf k'}^{\mu,\alpha}$ was used. The shifts in the momentum coordinates can now be written in terms of differential operators with the help of the identities introduced before together with $\mathbf q e^{i\mathbf q\cdot \mathbf r}  = -i \bm{\nabla}_\mathbf{r} e^{i\mathbf q\cdot \mathbf r}$, leading to
\begin{equation}
\mathcal C^{(1)}_\alpha   = \frac{Q_\alpha}{m_\alpha}\Bigg[\cos \left(\frac{1}{2} \overrightarrow{\bm \nabla}_\mathbf{r} \cdot \overrightarrow{\bm \nabla}_\mathbf{k}  \right) \mathbf  A(\mathbf r) \cdot\overrightarrow{\bm \nabla}_\mathbf{r}- \hbar\mathbf k  \cdot \mathbf A (\mathbf r)\frac{2}{\hbar}\sin \left(\frac{1}{2} \overleftarrow{\bm \nabla}_\mathbf{r}\cdot \overrightarrow{\bm \nabla}_\mathbf{k}  \right)  \Bigg]\mathcal W_\alpha(\mathbf r,\mathbf k). \label{CA1}
\end{equation}
\par 
A similar procedure provides
\begin{equation}
	\mathcal C^{(2)}_\alpha= \frac{i}{\hbar}\sum_{\mathbf q,\mu,\mu'}  \int d\mathbf r_1 \, \widetilde{D}^{\mu,\mu',\alpha}(\mathbf r_1-\mathbf r,\mathbf q) \exp\left(\frac{i}{2}\overrightarrow{\bm \nabla}_{\mathbf{r}_1} \cdot \overrightarrow{\bm \nabla}_\mathbf{q}   - \frac{i}{2}\overrightarrow{\bm \nabla}_{\mathbf{r}_1}\cdot  \overrightarrow{\bm \nabla}_\mathbf{k} \right)\mathcal N_{\mu ,\mu'}(\mathbf r_1,\mathbf q) \mathcal W_\alpha(\mathbf r,\mathbf k)+ \text{c.c.}  \label{C_2}
\end{equation} 
which, akin to Eq.~\eqref{Lambda2}, represents a linear operator acting on matter distributions and having a nonlocal dependence on the photon distribution. In the next section we show that the latter become local operators when the difference between the frequencies of light and plasma oscillations is sufficiently large. 

\subsection{High-frequency limit}\label{HFlimit}
For arbitrary distributions, the phase-space dynamics promoted by light-matter scattering events is represented by nonlocal collision operators of Eqs.~\eqref{Lambda2} and \eqref{C_2}. Therefore, density fluctuations at $\mathbf r$ couple to all values of  the distribution of scattering partners over the entire volume of the system, which translates the delocalization of both matter particles and photons through the space dependence of the Fourier transform $\widetilde{D}$ therein. \par 
When the difference between plasma and radiation length scales is sufficiently large, we expect locality to be recovered. This can be obtained by considering that the photon field is characterized by a central frequency $\omega$ that verifies $\omega \gg \omega_\text{p}$, with $\omega_\text{p}$ the plasma frequency. To second order in 
$\omega_\text{p}/\omega$, we find
\begin{equation}
	\widetilde{D}^{\mu,\mu',\alpha}(\mathbf r,\mathbf k) = \frac{\hbar Q_\alpha^2}{2 \epsilon_0 m_\alpha \omega_{\mathbf k}}\delta_{\mu,\mu'}\delta(\mathbf r)  + \order{\omega_\text{p}^2/\omega^2},\label{XmE}
\end{equation} 
where $\omega = ck$. By defining the local plasma frequency of $\alpha$ particles as
\begin{equation}
\Omega_{\alpha}(\mathbf r) = \sqrt{\frac{Q_\alpha^2 n_\alpha(\mathbf r)}{\epsilon_0 m_\alpha}}	,
\end{equation}
and neglecting $\order{\omega_\text{p}^2/\omega^2}$ terms, the photon scattering term becomes
\begin{align}
	\Lambda^{(2)}_{\mu,\mu'} &=  \frac{1}{\omega_\mathbf{k}} \sum_{\alpha} \Omega_\alpha^2(\mathbf r) \sin\left( \frac{1}{2} \overleftarrow{\bm{\nabla}}_{\mathbf k} \cdot  \overrightarrow{\bm{\nabla}}_{\mathbf r} - \frac{1}{2} \overleftarrow{\bm{\nabla}}_{\mathbf r} \cdot  \overrightarrow{\bm{\nabla}}_{\mathbf k}  \right) \mathcal N_{\mu,\mu'}(\mathbf r,\mathbf k),  \end{align} 	
which no longer couples different polarization elements. The matter counterpart becomes
\begin{align}
\mathcal C^{(2)}_\alpha	= \frac{Q_\alpha^2}{\epsilon_0 m_\alpha} \frac{1}{V} \sum_{\mathbf q} \mathcal N(\mathbf r,\mathbf q) \sin\left(\frac{1}{2}\overleftarrow{\bm \nabla}_{\mathbf{r}} \cdot \overrightarrow{\bm \nabla}_\mathbf{k}   + \frac{1}{2}\overleftarrow{\bm \nabla}_{\mathbf{r}}\cdot  \overrightarrow{\bm \nabla}_\mathbf{q} \right) \mathcal W_\alpha(\mathbf r,\mathbf k) \frac{1}{\omega_{\mathbf q}} ,
\end{align}
with $\mathcal N = \sum_{\mu} \mathcal N_{\mu,\mu}$ the trace of the photon distribution. \par 
It is worth noting that the off-diagonal terms of $\mathcal N_{\mu,\mu'}$ decouple when we make the high-frequency approximation. This means that polarization correlations due to scattering are only important when the photon and plasma length scales are comparable (or equivalently, when the energy of photons and plasma oscillations are comparable).  
\subsection{The semiclassical limit of $\mathcal C^{(1)}_\alpha$}\label{SAOC}
The goal of this section is to show that the linear matter collision term, defined by
\begin{equation}
\mathcal C^{(1)}_\alpha(\mathbf r,\mathbf p)   = \frac{Q_\alpha}{m_\alpha}\Bigg[\cos \left(\frac{\hbar}{2} \overrightarrow{\bm \nabla}_\mathbf{r} \cdot \overrightarrow{\bm \nabla}_\mathbf{p}  \right) \mathbf  A(\mathbf r) \cdot\overrightarrow{\bm \nabla}_\mathbf{r}\mathcal W_\alpha(\mathbf r,\mathbf p) - \mathbf p  \cdot \mathbf A (\mathbf r)\frac{2}{\hbar}\sin \left(\frac{\hbar}{2} \overleftarrow{\bm \nabla}_\mathbf{r}\cdot \overrightarrow{\bm \nabla}_\mathbf{p}  \right)  \mathcal W_\alpha(\mathbf r,\mathbf p)\Bigg],  \label{CA2}
\end{equation}
admits an $\hbar$-expansion of the form
\begin{equation}
	\mathcal C^{(1)}_\alpha(\mathbf r,\mathbf p) = -\mathbf  V_L(\mathbf r) \cdot \bm{\nabla}_{\mathbf r} \mathcal W(\mathbf r, \mathbf p)  - \mathbf F_L(\mathbf r,\mathbf p) \cdot\bm{\nabla}_{\mathbf p} \mathcal W(\mathbf r, \mathbf p) + \mathcal O(\hbar^2)  \label{ap3_2} 
\end{equation}
where $\mathbf  p = \hbar \mathbf k$ is the momentum, $\mathbf V_L(\mathbf r) = -Q_\alpha \mathbf A(\mathbf r)/m_\alpha$ is the diamagnetic velocity and
\begin{equation}
\mathbf F_L(\mathbf r,\mathbf p) = Q_\alpha \Big[\mathbf E(\mathbf r) + \frac{\mathbf p}{m_\alpha} \times \mathbf B(\mathbf r)\Big] \label{Lonrz1}
\end{equation}
is the Lorentz force. This expansion determines that the semiclassical limit of $\mathcal C^{(1)}_\alpha$ reduces to the classical interactions, since higher-order corrections contain at least second-order spatial derivatives. In what follows, we omit the index $\alpha$ to ease the notation.\par  
We start by replacing the differential operators in Eq.~\eqref{CA2} by their first Taylor component. The term containing the cosine immediately leads to the desired contribution involving the diamagnetic velocity. In order to arrive at the Lorentz force, we use Einstein's notation for repeated indices to write the second term as
\begin{align}
-\frac{Q}{m}\mathbf p  \cdot \mathbf A (\mathbf r)\overleftarrow{\bm \nabla}_\mathbf{r}\cdot \overrightarrow{\bm \nabla}_\mathbf{p}   \mathcal W(\mathbf r,\mathbf p) = -\frac{Q}{m} p_\ell\frac{\partial A_\ell(\mathbf r)}{\partial r_j} \frac{\partial \mathcal W(\mathbf r,\mathbf p)}{\partial p_j} . \label{ap3_3}
\end{align} 
Let us now evaluate the $j$ component of $\mathbf p \times \mathbf B$, 
\begin{align*}
 (\mathbf p \times \mathbf B	)_j &=  (\mathbf p \times \bm{\nabla}\times \mathbf A)_j \\
 &= \varepsilon_{jk\ell} \, p_k (\bm{\nabla}\times \mathbf A)_\ell \\
 &= \varepsilon_{jk\ell} \, \varepsilon_{m n \ell} p_k \frac{\partial A_n}{\partial r_m}.
\end{align*}
Using the property $\varepsilon_{jk\ell} \, \varepsilon_{m n \ell}  = \delta_{j,m} \delta_{k,n} - \delta_{j,n}\delta_{k,m}$ we obtain
\begin{align}
 (\mathbf p \times \mathbf B	)_j &=  p_\ell \frac{\partial A_\ell}{\partial r_j} - p_\ell \frac{\partial A_j}{\partial r_\ell}, 
\end{align}
such that Eq.~\eqref{ap3_3} becomes
\begin{align}
-\frac{Q}{m}\mathbf p  \cdot \mathbf A (\mathbf r)\overleftarrow{\bm \nabla}_\mathbf{r}\cdot \overrightarrow{\bm \nabla}_\mathbf{p}   \mathcal W(\mathbf r,\mathbf p) =  -Q\frac{p_\ell}{m} \frac{\partial A_j}{\partial r_\ell}\frac{\partial \mathcal W(\mathbf r,\mathbf p)}{\partial p_j} -  Q\Big[\frac{\mathbf p}{m} \times \mathbf B(\mathbf r)	\Big] \cdot \bm{\nabla}_{\mathbf p}  \mathcal W(\mathbf r,\mathbf p) . \label{LRNZT}
\end{align} 
Because the last term is already the magnetic contribution to $\mathbf F_L$, then the first term must be the electric counterpart. If this is the case, then the factor $p_\ell$ multiplied by the spatial derivative acting on $A_j$ must somehow be replaced by a partial time derivative, since we want $\mathbf E = -\partial_t \mathbf A$ to appear. However, one easily concludes that this substitution is not exact. In fact, Eq.~\eqref{LRNZT} defines a modified Lorentz force $\widetilde{\mathbf F}_L$ of the form
\begin{equation}
	\widetilde{\mathbf F}_L(\mathbf r,\mathbf p) = \frac{Q}{m}\Big[ (\mathbf p \cdot \bm{\nabla}) \mathbf A(\mathbf r) + \mathbf p \times \mathbf B(\mathbf r)\Big], 
\end{equation} 
which reduces to Eq.~\eqref{Lonrz1} when $\hbar$ corrections are neglected. We will now show that $\widetilde{\mathbf F}_L = \mathbf F_L + \mathcal O(\hbar^2)$, such that the force term in Eq.~\eqref{ap3_2} is indeed the desired classical expression. \par 
Let us recall the relation between $\mathcal W(\mathbf r,\mathbf p)$ and its classical counterpart $ f(\mathbf r,\mathbf p)$. Since quantum contributions to the Wigner equation are at least of order $\hbar^2$ (see, e.g., Ref.~\cite{moyal}), this means that 
\begin{equation}
	\mathcal W(\mathbf r,\mathbf p) = f(\mathbf r,\mathbf p) + \mathcal O(\hbar^2).
\end{equation}
Therefore, for the desired accuracy, it is sufficient to replace $\mathcal W$ by $f$ when evaluating the first term on the right-hand side of Eq.~\eqref{LRNZT}. Given that $f$ represents a classical ensemble of $N$ particles moving in classical trajectories $\bm x_s(t)$ and $\bm q_s(t)$ which verify Hamilton's equations, then a solution is
\begin{equation}
	f(\mathbf r,\mathbf p)= \sum_{s=1}^N \delta(\mathbf  r-\bm x_s(t))\delta(\mathbf  p-\bm q_s(t)).
\end{equation}
We get
\begin{align}
-Q\frac{p_\ell}{m} \frac{\partial A_j(\mathbf r,t)}{\partial r_\ell}\frac{\partial f(\mathbf r,\mathbf p)}{\partial p_j}  &= -Q\frac{\partial}{\partial p_j} \frac{p_\ell}{m} \frac{\partial A_j(\mathbf r)}{\partial r_\ell} f(\mathbf r,\mathbf p), \nonumber \\
&= - Q \frac{\partial}{\partial p_j} \sum_{s=1}^N  \frac{p_\ell}{m} \frac{\partial A_j(\mathbf r,t)}{\partial r_\ell} \delta(\mathbf  r-\bm x_s(t))\delta(\mathbf  p-\bm q_s(t)), \nonumber \\
& = - Q \frac{\partial}{\partial p_j} \sum_{s=1}^N  \frac{q_{s,\ell}(t)}{m} \frac{\partial A_j(\mathbf r,t)}{\partial r_\ell} \delta(\mathbf  r-\bm x_s(t))\delta(\mathbf  p-\bm q_s(t)), \nonumber \\
 &= - Q \frac{\partial}{\partial p_j} \sum_{s=1}^N  \frac{d x_{s,\ell}(t)}{d t} \frac{\partial A_j(\bm x_s(t),t)}{\partial  x_{s,\ell}(t)} \delta(\mathbf  r-\bm x_s(t))\delta(\mathbf  p-\bm q_s(t)).\nonumber 
\end{align} 
In the first equality we used $\bm{\nabla}\cdot \mathbf A=0$, while the remaining steps follow from standard properties of the delta function together with Hamilton's equations. The electric-field components can now be introduced through the relation
\begin{equation}
	\frac{d x_{s,\ell}(t)}{d t} \frac{\partial A_j(\bm x_s(t),t)}{\partial  x_{s,\ell}(t)}  =  \frac{d}{dt}  A_j(\bm x_s(t),t) - \frac{\partial }{\partial t} A_j(\bm x_s(t),t) = \frac{d}{dt}  A_j(\bm x_s(t),t) + E_j(\bm x_s(t),t),
\end{equation} 
 where $d/dt$ denotes the total time derivative. Finally, we take advantage of gauge freedom and apply a gauge transformation $\mathbf A \rightarrow \mathbf A'  = \mathbf A + \bm{\nabla}\chi $ so that $d\mathbf A'/dt = 0$,  which leads to
 \begin{align*}
-Q\frac{p_\ell}{m} \frac{\partial A_j(\mathbf r,t)}{\partial r_\ell}\frac{\partial f(\mathbf r,\mathbf p)}{\partial p_j}  &=  - Q \frac{\partial}{\partial p_j} \sum_{s=1}^N  E_j(\bm x_s(t),t)\delta(\mathbf  r-\bm x_s(t))\delta(\mathbf  p-\bm q_s(t)), \\
&= -Q E_j(\mathbf r,t) \frac{\partial}{\partial p_j} \sum_{s=1}^N \delta(\mathbf  r-\bm x_s(t))\delta(\mathbf  p-\bm q_s(t)), \\
&= -Q \mathbf E(\mathbf  r,t)  \cdot \bm{\nabla}_{\mathbf p}  f(\mathbf r,\mathbf p), 
\end{align*} 
as intended. 
 \end{widetext}
\section{The classical photon distribution}\label{apB}
Since the advent of quantum kinetics, many authors have applied the Wigner theory to the study of classical electrodynamical systems \cite{phW0,phW1,phW2,phW3}. In these works, the unquantized electromagnetic field is replaced by a quasi-distribution function defined by
\begin{equation}
\mathcal N_\text{cl.}(\mathbf r,\mathbf k,t) = y_{\mathbf k}\int d\mathbf r' e^{i\mathbf k\cdot \mathbf r'} \mathbf{E}(\mathbf r_-,t)\cdot \mathbf{E}^\ast(\mathbf r_+,t), \label{ClassWPh}
\end{equation}
where $\mathbf{E}$ is the electric field and $\mathbf r_\pm = \mathbf r \pm \mathbf r'/2$. The coefficients $y_{\mathbf k}$ are determined from the solutions of the free Maxwell's equation that verify the boundary conditions of the problem, after imposing 
\begin{equation}
I_\text{cl.}(\mathbf r,t) = \frac{1}{V} \sum_{\mathbf k} g_s\hbar\omega_{\mathbf k}c \, \mathcal N_\text{cl.}(\mathbf r,\mathbf k,t) ,\label{Iclass}
\end{equation} 
with $I_\text{cl.}= \epsilon_0 c|\mathbf{E}|^2$ being the classical intensity and $g_s=2$ the spin degeneracy. For a bulk plasma we get $y_{\mathbf k} = \epsilon_0/(g_s \hbar\omega_{\mathbf k})$. \par 
Equation~\eqref{Iclass} can be interpreted as the energy-current density of an ensemble of particles described by the classical distribution $\mathcal N_\text{cl.}$. Moreover, one can show that $n(\mathbf r) = V^{-1} \sum_{\mathbf k} \mathcal N_\text{cl.}(\mathbf r,\mathbf k)$ and $\widetilde{n}(\mathbf k) = V^{-1}\int d\mathbf r \, \mathcal N_\text{cl.}(\mathbf r,\mathbf k)$ are positive-definite quantities and reduce, respectively, to the photon density and occupation number that one expects from classical electromagnetic arguments.\par
In the main text we use a different photon distribution, Eq.~\eqref{PWignerF2}, which fully accounts for photon fluctuations and reduces to Eq.~\eqref{ClassWPh} when fluctuations are neglected. For classical states of light, the description in terms of fields or distribution functions are equivalent due to the one-to-one correspondence between $\mathbf{E}$ and $\mathcal N_\text{cl.}$. On the other hand, for quantum states of light with large photon fluctuations, there will be a fluctuating contribution to the distribution function defined by $\Delta\mathcal{N} \equiv \mathcal N - \mathcal N_\text{cl.}$, which is related to field fluctuations as
\begin{equation}
 \Delta\mathcal{N} (\mathbf r,\mathbf k,t) =  \sum_{\mathbf  q,\mu}  e^{i\mathbf q\cdot \mathbf r} \langle \delta\hat a^\dagger_{\mathbf k - \mathbf q/2,\mu} \delta\hat a_{\mathbf k + \mathbf q/2,\mu} \rangle .
\end{equation}
When $\Delta\mathcal{N}$ is sufficiently large, using the classical distribution of Eq.~\eqref{ClassWPh} no longer leads to a complete description, and Eq.~\eqref{PWignerF2} must be used. Note that, due to fluctuations, the total photon distribution $\mathcal N_\text{cl.} + \Delta \mathcal N$ is no longer related to the classical fields and the two must, instead, be treated as independent variables that are dynamically coupled through Eqs.~\eqref{MXWeq} and \eqref{photon1} of the main text. \par 
In order to clarify the difference between the two approaches, let us calculate the equation of motion for $\mathcal N_\text{cl.}$, which follows from Maxwell's equation
\begin{equation}
	\frac{\partial}{\partial t} \mathbf{E}(\mathbf r,t) = c\bm\nabla\times \mathbf B(\mathbf r,t) -\frac{1}{\epsilon_0} \bm j (\mathbf r,t),
\end{equation}
and compare it with Eq.~\eqref{photon1}. After standard manipulations, we find 
\begin{align}
&	\Bigg[ \frac{\partial}{\partial t} + 2\omega_{\mathbf k} \sin\Big(\frac{1}{2} \overleftarrow{\bm{\nabla}}_\mathbf{k} \cdot \overrightarrow{\bm{\nabla}}_\mathbf{r} \Big) \Bigg]  \mathcal N_\text{cl.}\nonumber\\
	 &=-\frac{1}{\hbar\omega_{\mathbf k}} \int d\mathbf r' \ e^{i\mathbf k\cdot\mathbf r'} \bm{j}(\mathbf r_{+},t)\cdot \mathbf E^\ast(\mathbf r_{-},t)   + \text{c.c}. 
\end{align} 
Note that the right-hand side corresponds to the classical contribution of Eq.~\eqref{lambda1}, i.e., taking $\overleftrightarrow{\mathcal U} = \mathbb{1}$. While the free dynamics of the classical distribution converges to its quantum counterpart, this is not the case for the light-matter collision terms, where field fluctuations become significant. Specifically, all nonlinear scattering terms coupling matter and light distributions are absent, along with the higher-order corrections represented by $\Lambda^{(1)}$. These corrections are linked to photon-polarization dynamics, which is neglected by the classical photon distribution.

\bibliographystyle{apsrev4-2}
\bibliography{references.bib}

\end{document}